\documentclass[preprint,showpacs,preprintnumbers,amsmath,amssymb,nofootinbib]{revtex4}
\pdfoutput=1

\usepackage{etex}

\usepackage{amsthm,amscd,amsbsy,array}
\usepackage{bm}% bold math
\usepackage{soul} % underline, strikethrough, etc.

\usepackage[utf8]{inputenc}
\usepackage[T1]{fontenc}

\usepackage{graphics,graphicx,xcolor}

\usepackage{soul} % underline, strikethrough, 

\numberwithin{equation}{section}

\usepackage[colorlinks=true, pdfstartview=FitV, linkcolor=blue, citecolor=blue, urlcolor=blue]{hyperref} % hyperref

\newcommand{\magenta}{\textcolor{magenta}}

\newcommand{\blue}{\textcolor{blue}}

\newcommand{\gb}{\colorbox{green}}

\newcommand{\dgreen}{\textcolor[rgb]{0,0.35,0}}

\newenvironment{redtext}{\color{red}}{\ignorespacesafterend} 
\newenvironment{bluetext}{\color{blue}}{\ignorespacesafterend} 
\newenvironment{greentext}{\color{green}}{\ignorespacesafterend}
\newenvironment{dgreentext}{\color{\dgreen}}{\ignorespacesafterend}
 
\newcommand{\bblue}{\begin{bluetext}} 
\newcommand{\eblue}{\end{bluetext}} 
\newcommand{\bred}{\begin{redtext}}
\newcommand{\ered}{\end{redtext}}
\newcommand{\bgreen}{\begin{greentext}}
\newcommand{\egreen}{\end{greentext}}
\newcommand{\bdgreen}{\begin{dgreentext}}
\newcommand{\edgreen}{\end{dgreentext}}
\numberwithin{equation}{section}

\let\ssection=\section
\renewcommand{\section}{\setcounter{equation}{0}\ssection}

\newcommand{\cA}{{\mathcal{A}_{+}}}

\newcommand{\bb}{{\bf b}}

\newcommand{\bbeta}{\boldsymbol{\beta}}

\newcommand{\cC}{{\mathcal{C}}}

\newcommand{\cD}{{\mathcal{D}}}

\newcommand{\diag}{\mathrm{diag}}

\newcommand{\cH}{{\mathcal{H}}}

\newcommand{\cK}{{\mathcal{K}}}

\newcommand{\cL}{{\mathcal{L}}}

\newcommand{\cM}{\mathcal{M}}

\newcommand{\bP}{{\bf P}}

\newcommand{\cQ}{\mathcal{Q}}
\newcommand{\cT}{\mathcal{T}}

\newcommand{\bx}{{\bm{x}}}

\newcommand{\cR}{\mathcal{R}}

\newcommand{\Tr}{\mathrm{Tr}}

\newcommand{\bW}{{\bf W}}

\newcommand{\bX}{{\bm X}}

\def\smallover#1/#2{\hbox{$\textstyle\frac{#1}{#2}$}} %

\def\where{{\quad\text{where}\quad}}
\def\with{{\quad\text{with}\quad}}
\def\aand{{\quad\text{and}\quad}}
\def\Ort{{\rm O}}

\def\bP{{\bm{P}}}

\def\vX{\mathbf{X}}
\def\parag{\hfil\break} %%%%% paragraph
\def\kikezd{\parag\underbar}

\def\benu{\begin{enumerate}}
\def\eenu{\end{enumerate}}
\def\beq{\begin{equation}}
\def\eeq{\end{equation}}
\def\beqa{\begin{eqnarray}}
\def\eeqa{\end{eqnarray}}
\def\nn{\nonumber}
\def\barray{\left(\begin{array}}
\def\earray{\end{array}\right)}
\def\barraynb{\begin{array}}
\def\earraynb{\end{array}}

\def\Ort{{\rm O}}
\def\IR{{\mathbb{R}}} %%%%% Reals

\def\?{\quad{\gb{\fbox{\texttt{?}}\;}}\quad}
\def\p{{\partial}}

\def\beq{\begin{equation}}
\def\eeq{\end{equation}}
\def\bea{\begin{eqnarray}}
\def\eea{\end{eqnarray}}

\def\p{\partial}

\def \p{{\partial}}

\def\bP{\mathbb P}

\def\6{\partial}
\def\7{\tilde}
\def\8{\widehat}
 \def\bx{{\bf x}}

\def\pa{\partial}

\def\bP{{\rm{\bf P}}}

\def\G11{\Gamma_{11} }

\newcommand{\const}{\mathop{\rm const.}\nolimits}
\newcommand{\half }{\frac{1}{2}}

\def\smallover#1/#2{\hbox{$\textstyle\frac{#1}{#2}$}} %
\def\smallcirc{{\raise 0.5pt \hbox{$\scriptstyle\circ$}}}
\def\2{{\smallover1/2}}

\newcommand{\bigbox}[1]{\fbox{%
\rule[-20pt]{0pt}{45pt}$\;\;\displaystyle{#1}\;\;$}
}
\newcommand{\medbox}[1]{\fbox{%
\rule[-10pt]{0pt}{25pt}$\;\;\displaystyle{#1}\;\;$}%
}

\let\ssection=\section
\renewcommand{\section}{\setcounter{equation}{0}\ssection}

\def\besub{\begin{subequations}}
\def\esub{\end{subequations}}
\begin{document} 

\preprint{%\texttt{arXiv:1905.08661v5  [gr-qc]}
}

\title{Scaling and conformal symmetries \\
for  plane gravitational waves
\\[6pt]
}

\author{
P.-M. Zhang${}^{1,2}$\footnote{e-mail: zhangpm5@mail.sysu.edu.cn},
M. Cariglia${}^{3}$\footnote
{e-mail: marco.cariglia@gmail.com},
M. Elbistan${}^{2,4}$\footnote{mailto: mahmut.Elbistan@lmpt.univ-tours.fr},
P. A. Horvathy${}^{2,4}$\footnote{mailto:horvathy@lmpt.univ-tours.fr}
}

\affiliation{
${}^1$ School of Physics and Astronomy, Sun Yat-sen University, Zhuhai, China
\\
${}^2$Institute of Modern Physics, Chinese Academy of Sciences
\\ Lanzhou, China
\\
${}^{3}${\small DEFIS, Universidade Federal de Ouro Preto,  MG-Brasil},
\\
${}^4$ Institut Denis Poisson CNRS/UMR 7013 - Universit\'e de Tours - Universit\'e d'Orl\'eans Parc de Grandmont, 37200, Tours, (France)
}

\date{\today}


\begin{abstract} 

The isometries of an exact plane gravitational wave are symmetries for both massive and massless particles. Their conformal extensions are in fact chrono-projective transformations (introduced earlier by Duval et al), and are symmetries for massless particles. Homotheties are universal chrono-projective symmetries for any profile. Chrono-projective transformations also generate new conserved quantities for the underlying non-relativistic systems in the Bargmann framework.
Homotheties play a similar role for the lightlike ``vertical'' coordinate as isometries play for the transverse coordinates.  
\\[4pt]
\noindent
J. Math. Phys. 61, 022502 (2020); https://doi.org/10.1063/1.5136078\\ 
\texttt{arXiv:1905.08661v5 [gr-qc]}
\end{abstract}

\maketitle

%\tableofcontents

\newpage

%%%%%%%%%%%%%%%%%%%%%%%%%%%%%%%%%%%%%%%%%%%%%%%%%%%%%%%%%%%%%%%%%%%%%%%%%%%%%%
%%%%%%%%%%%%%%%%%%%%%%%%%%%%%%%%%%%%%%%%%%%%%%%%%%%%%%%%%%%%%%%%%%%%%%%%%%%%%%
\section{Introduction}\label{Intro}
%%%%%%%%%%%%%%%%%%%%%%%%%%%%%%%%%%%%%%%%%%%%%%%%%%%%%%%%%%%%%%%%%%%%%%%%%%%%%%
%%%%%%%%%%%%%%%%%%%%%%%%%%%%%%%%%%%%%%%%%%%%%%%%%%%%%%%%%%%%%%%%%%%%%%%%

A $4-$dimensional plane wave spacetime is given, in Brinkmann coordinates \cite{Brinkmann}, by
\beq
g_{\mu\nu}dX^\mu dX^\nu=\delta_{ij} dX^i dX^j + 2 dU dV + K_{ij}(U) X^i X^j dU^2 \,,
\label{Bmetric}
\eeq
where $\vX\in \IR^2$ and $U,\, V$ are the transversal resp. light-cone coordinates. (\ref{Bmetric}) admits a covariantly constant, null Killing vector $\xi = \partial_V$ and it has a symmetric profile $K_{ij}(U)$.  If, in addition, $K_{ij}$ is traceless then the elements of the Ricci tensor vanish identically,  $R_{\mu\nu}=0$, and (\ref{Bmetric}) satisfy vacuum  Einstein equations, i.e., it is an \emph{exact plane gravitational wave} \cite{BoPiRo}.

Recent insight into the  ``Memory Effect'' for gravitational waves \cite{Memory,OurMemory,Harte,Faber,Shore,Maluf,Kulczycki, POLPER,Ilderton}
was brought about by a better understanding of their symmetries. For exact plane waves (\ref{Bmetric}) the implicitly known isometry group \cite{BoPiRo,EhlersKundt,Sou73,exactsol,Torre} maps geodesics to geodesics  and  yields conserved quantities~;  conversely, the latter determine the transverse motion (motion in the $\vX$-plane) of test particles in the gravitational wave background \cite{Sou73,Carroll4GW}. 

The isometries of plane gravitational waves (\ref{Bmetric}) have been identified as L\'evy-Leblond's ``Carroll'' group  with broken rotations \cite{Leblond,Carroll4GW,Carrollvs,NewCarroll,Morand:2018tke,Ciambelli:2019lap,SLC}. 
However  \emph{homotheties}, $h~:~\cM \to \cM$\,,
\beq
U \to U,
\quad
\bX \to \chi\, \bX,
\quad
V \to \chi^2\, V,
\quad
\chi\,=\const
\label{homothety0}
\eeq
play also an important role, namely for the integrability of the geodesic equations \cite{AndrPrenc, AndrPrenc2, PKKA}.
The homothety is  \emph{not} an isometry though ; it is  a \emph{conformal transformation}, i.e., infinitesimally 
\beq
L_Yg_{\mu\nu}=2{\omega}g_{\mu\nu}\,  
\label{confodef0}
\eeq
for some function $\omega$.   For the homothety  (\ref{homothety0}) $\omega= 1$.
Finding all conformal symmetries of (\ref{Bmetric}) is a difficult task which requires a series of constraints to be satisfied whose solution depends on the chosen profile  and is found only case-by-case \cite{Sippel, Eardley, MaMa,Keane, HallSteele,KeaTu04}.

On the other hand, plane gravitational waves endowed with a covariantly constant null vector $\xi$ (\ref{Bmetric}) can also be viewed as a ``Bargmann manifold'' for a non-relativistic system in one less dimensions. 
The underlying non-relativistic motions can be ``Eisenhart-Duval (E-D) lifted'' as null geodesics \cite{Eisenhart,Bargmann,DGH91,dissip}. The Bargmann point of view   provides a powerful framework to investigate the symmetries of the associated non-relativistic system.

Each conformal vector field, (\ref{confodef0}), of the metric generates a conserved quantity $\cQ$ for null geodesics. 
If $Y$ preserves, in addition, the vertical vector $\xi$, 
\beq
 L_Y\xi=0,
 \label{Lxi0}
\eeq
 then $\cQ$  projects to a conserved quantity for the underlying non-relativistic dynamics. Conformal vector fields which satisfy also (\ref{Lxi0}) generate the ``extended Schr\"odinger group'' ; such isometries span the  ``Bargmann group'' \cite{Bargmann,DGH91}.
 
Up in the Bargmann space a conserved quantity 
is associated with any conformal vector field, even if the latter does not preserve $\xi$, though.  For instance,  the generator $Y_h=(Y_h^\mu)$ of the homothety  (\ref{homothety0})  preserves only the \emph{direction} of the vertical vector 
\beq
L_{Y_h}\xi=\psi\,\xi,
\qquad
\psi=-2\chi=\const\, 
\label{Homocond}
\eeq
and generates the charge \eqref{Qhomot} whose conservation determines the vertical coordinates, see sec.  \ref{homoSec} below. 

The first relation here is in fact a paradigm of the ``chrono-projective condition''  
\beq
L_Y\xi=\psi\,\xi\,\qquad \psi= \const
\label{Chronocond}
\eeq
which plays a fundamental r\^ole in our investigations. 
 The constant $\psi$ is called the chrono-projective factor.

This condition has been considered at various instances. 

Firstly, it was put forward  by Duval et al. \cite{DGH91, 5Chrono}. Remarkably, these authors introduced it originally as a geometric property related to the Newton-Cartan structure of 
$d+1$-dimensional \emph{non-relativistic spacetime}. They called it ``chrono-projective property''. Later these same authors realized that chrono-projectivity can actually be derived by lightlike reduction from a $d+1,1$ dimensional \emph{relativistic} spacetime --- their  Bargmann space \cite{5Chrono,DuLaz}, --- namely as in \eqref{Chronocond}. See also eqn.  \# (4.4) of \cite{DGH91} or \# (5.17)-(5.21) of \cite{DuLaz}. In \cite{NCosmo} it was  rebaptized as the \emph{conformal Newton-Cartan} group ; in \cite{Gundry} it was rediscovered under the name of  ``enlarged Schr\"odinger group''. In this paper we return to the original terminology proposed in \cite{DThese}.
 
The condition (\ref{Chronocond}) is \emph{almost} identical to a property noticed  by Hall et al \cite{HallSteele}, who pointed out that  for pp waves it follows from studying the Weyl tensor -- however with $\psi$ a \emph{function}, rather then a \emph{constant}.
The original definition made in \cite{5Chrono} would allow $\psi$ to be a \emph{function}. However, the additional condition \# (4.8)  these authors impose on the connection $\Gamma$ implies  that $\psi$ is necessarily a \emph{constant}. It is (\ref{Chronocond}) (and not the formula of \cite{HallSteele}) that reproduces the chrono-projectivity after reduction.
%%%%

In sec.\ref{ConfChronSec} we show that  \emph{all special conformal Killing vectors of a non-flat pp-wave (\ref{Bmetric}) are chrono-projective.} Thus all conformal Killing vector of an Einstein vacuum solution satisfies (\ref{Chronocond}). 
 
 In this paper we take, on the one hand, advantage of the chronoprojective condition (\ref{Chronocond}) to simplify the procedure of finding all conformal vectors  of gravitational waves and derive, on the other hand, the associated conserved charges. 

Our paper is organized as follows: In Sec.\ref{planewaves}, after  recalling exact plane waves and conformal transformations, we briefly outline  the Bargmann [alias Eisenhart-Duval] approach. Chrono-projective transformations are then introduced.
Conformal transformations  and their subgroups  in flat space  are spelled out in sec. \ref{MinkowskiSec}.
In sec. \ref{JacobiSec}, conserved quantities of null geodesics related to conformal transformations are discussed. %\footnote{
(Symmetries of timelike geodesics were considered recently using non-local conservation laws \cite{Dimakis}).
%}. 
In sec. \ref{homoSec}, new types of conserved quantities associated with chrono-projective transformations in the Bargmann  framework,
generalizing those in \cite{KHarmonies} are considered.  
The chrono-projective transformations  of exact plane gravitational waves are studied in sec.\ref{BJRSec}, 
using Baldwin-Jeffery-Rosen (BJR) \cite{BaJe} coordinates.
 In sec.\ref{Examples} we illustrate our general theory  on various examples. 
\goodbreak

%%%%%%%%%%%%%%%%%%%%%%%%%%%%%%%%%%%%%%%%%%%
\section{Exact plane gravitational waves}\label{planewaves}
%%%%%%%%%%%%%%%%%%%%%%%%%%%%%%%%%%%%%%%%%%%

%%%%%%%%%%%%%%%%%%%%%%%%%%%%%%%%%%%%%%%%%%%
\subsection{Gravitational waves and conformal transformations}\label{GWConfSec}
%%%%%%%%%%%%%%%%%%%%%%%%%%%%%%%%%%%%%%%%%%%

For generic profile $K_{ij}(U)$,  the isometries of an exact gravitational plane wave (\ref{Bmetric}) i.e. diffeomorphisms  of spacetime, $f:~\cM \to \cM$ s.t.
\beq
f^*g_{\mu\nu}= g_{\mu\nu} \quad\text{\small infinitesimally}\quad
L_Yg_{\mu\nu}=0
\label{isom}
\eeq 
span a  $5$-parameter group \cite{BoPiRo,EhlersKundt,Sou73,Torre,exactsol}, 
which is in fact the subgroup of the Carroll group in $2+1$ dimensions 
with broken rotations \cite{Leblond,Sou73,Carrollvs,NewCarroll,Carroll4GW}.
However the homothety (\ref{homothety0}) \cite{Torre,AndrPrenc} 
generated by the vector field
\beq
Y_{hom}= X^i\p_{i}+2 V\p_{V}\,
\label{infhomo}
\eeq
is \emph{not} an isometry but
 a \textit{conformal transformation}  of the pp-wave metric (\ref{Bmetric}), 
% \footnote{I.e., a diffeomorphism under which the metric pulls back to an (in general position dependent) positive multiple of itself.  
%This should be distinguished from a ``Weyl rescaling" of the metric by which the metric is replaced by a positive, in general position dependent, multiple of itself at the same point, i.e. in the same coordinate system.},  
\beq
f^*g_{\mu\nu}= \Omega^2g_{\mu\nu} 
\qquad\text{\small infinitesimally}\qquad
L_Yg_{\mu\nu}=2\omega\,g_{\mu\nu}\,.
\label{confom}
\eeq 

For the homothety (\ref{homothety0})  
$\Omega^2=\chi^2=\const$. Its role may be  understood by looking at the geodesic motion.
For the profile 
\beq
K_{ij}(U){X^i}{X^j}=
\half{\cA}(U)\Big((X^1)^2-(X^2)^2\Big)+{\mathcal{A}}_{\times}(U)\,X^1X^2\,,
\label{Bprofile}
\eeq
where $\cA$ and ${\mathcal{A}}_{\times}$ are the $+$ and $\times$ polarization-state amplitudes \cite{Brinkmann,BoPiRo,EhlersKundt,exactsol}, the geodesics are described by
\begin{subequations}
\begin{align}
& \dfrac{d^2 \bX}{dU^2} - \half\barray{lr}
{\cA} &{\mathcal{A}}_{\times}
\\
{\mathcal{A}}_{\times} & -{\cA}
\earray
\bX = 0\,,
\label{ABXeq}
\\[10pt]
& \dfrac {d^2 V}{dU^2} + \dfrac{1}{4} \dfrac{d{\cA}}{dU\,}\Big((X^1)^2 - (X^2)^2 \Big)
+{\cA}\Big(X^1\dfrac{dX^1}{dU\,} - X^2\dfrac{dX^2}{dU\,}\Big)
\nn
\\[6pt]
&\qquad\,+\dfrac{1}{2} \dfrac {d{\mathcal A}_{\times}}{dU\,} X^1 X^2
+ {{\mathcal A}_{\times}}\Big(X^2\dfrac{dX^1}{dU\,} + X^1\dfrac{dX^2}{dU\,}\Big)
= 0\,.
\label{ABVeq}
\end{align}
\label{ABeqs}
\end{subequations}
Then the homothety (\ref{homothety0})  multiplies the $\bX$~-~equation by $\chi$ and the $V$-equation by $\chi^2$~; trajectories are therefore taken  to trajectories, as illustrated on fig.\ref{blowup}. 

Alternatively, the geodesic Lagrangian
\beq
\cL_{geo}=\half\delta_{ij} \dot{X}^i \dot{X}^j + \dot U \dot V + \half K_{ij}(U) X^i X^j \dot U^2 \,,
\label{geoLagrangian}
\eeq
where the dot means derivation w.r.t. an affine parameter $\sigma$ \footnote{We  mostly choose  $\dot{\{\,\cdot\,\}}= d/dU$.} scales under as, 
\beq
\cL_{geo} \to \chi^2\, \cL_{geo}\,,
\label{Lscale}
\eeq 
implying again  that the trajectories go into trajectories~:
The geodesic motion in such a background is  thus \emph{scale invariant}. We note that  all of these $4D$ trajectories project  to the same curve $\bX(U)$ in the transverse plane. Let us record for further use that 
\beq
\cL_{geo}=0
\eeq
for  \emph{null geodesics}, which are thus homothety-invariant  by (\ref{Lscale}).

We note that the transverse  equations (\ref{ABXeq}) are decoupled from the ``vertical'' one, (\ref{ABVeq}),  and can be solved separately. Once $\bX(U)$ has been determined, the result should be inserted into (\ref{ABVeq}) which then can be integrated. Analytic solutions are difficult to find, and therefore the best is to use numerical integration  \cite{POLPER}. As it will be further discussed in sec.\ref{homoSec}  the ``new'' conserved charge $\cQ_{hom}$ associated with the homothety provides an alternative way to derive the vertical motion.

For further use, we  record some basic facts about the conformal transformations (\ref{confom})~: for a conformally flat spacetime, the number of conformal transformations is 15, 
but for a non-conformally-flat spacetime, their maximum number  is  7 \cite{HallSteele}. 5 of them are isometries  and there is 1 homothety. There may or may not be a 7th transformation which may or may not be an isometry depending on special conditions, see sec. \ref{Examples}.

%%%%%%%%%%%%%%%%%%%%%%%%%%%%%%%%%%%%%%%%%%%%%%%%
\subsection{All conformal Killing vectors of vacuum pp-waves are chrono-projective
}\label{ConfChronSec}
%%%%%%%%%%%%%%%%%%%%%%%%%%%%%%%%%%%%%%%%%%%%%%%%%

The aim of this subsection is to prove the theorem stated in the title.

In their seminal paper Maartens and Maharaj  \cite{MaMa} have shown that for a pp-wave 
the conformal factor of the most general conformal Killing vector $\bm{Y}$ is (cf. their Eqns.\#  (29-32))
\beq
\omega (U, X, V) = \mu V + a'_i(U) X^i + b(U)\,,
\label{MaMaomega}
\eeq
where the  constant $\mu$ and the functions $a_i(U)$ and $b(U)$  can be determined by a case-by-case calculation using the additional constraints.
Now, 
\beq
L_{\bm{Y}} \partial_V = - \Big[ 2\mu V + a'_i(U) X^i + 2b(U) -a'(U)\Big] \partial_V - \Big[\mu X^i +  a_i (U) \Big]\partial_i\,, 
\label{MaMaLie}
\eeq
which does not seem to be parallel to $\xi=\p_V$. Accordingly,  the conformal factor (\ref{MaMaomega}) depends on all coordinates. However, in order to conclude whether $\bm{Y}$ is chrono-projective or not, one needs to deal with profile-dependent integrability conditions. Taking into account the additional integrability constraints, Maharaj and Maartens found, after tedious calculations, that for a \emph{special conformal Killing vector} $\bW$ in a non-flat pp-wave. (A conformal Killing vector is called \emph{special}  when its conformal factor
satisfies $\omega_{;\mu\nu}=0$.) 
\besub
\begin{align}
\bm{W} &= \rho (U^2 \partial_U + \frac{1}{2}\delta_{ij}X^i X^j \partial_V + U X^i\partial_i ) + \bm{Z}
\label{MaMaW}
\\
\label{mamaHom}
\bm{Z} &= \phi(2V \partial_V + X^i\partial_i) + \bm{X},
\\
\label{mamaKilling}
\bm{X} &= (\alpha U + \beta)\partial_U + (\lambda - \alpha V + c'_i(U)X^i )\partial_V + (c_i + \gamma \epsilon_{ij}X^j)\partial_i
\end{align}
\esub
where $\rho, \phi, \alpha, \beta, \gamma$ are constants and and $c_i(U)$ is a function, cf. their eqn. $\#$ (56). The first term in $\bm{Z}$ is a homothety ; $\bX$ is a Killing vector.
An  additional integrability condition, their eqn. \# (55), should also be satisfied. 

\emph{The special conformal Killing vector $\bm{W}$ is chrono-projective}, with conformal and chrono-projective factors 
\beq
\omega = \omega(U) = \rho U + \phi \;\aand\;  \psi=\alpha -2\phi\,,
\eeq
respectively. To complete the proof it is enough to remember that \emph{every conformal Killing vector of an Einstein vacuum pp-wave is  special conformal} \cite{MaMa}.  
\goodbreak

In an alternative approach inspired by \cite{HallSteele}, one starts with a conformal vector field $Y^\mu, L_Y g_{\mu\nu} = 2\omega g_{\mu\nu}
$, which satisfies
$L_Y C^\mu_{\nu\rho\sigma} =0$ , where $C^\mu_{\nu\rho\sigma}$ is the Weyl tensor.
 $k^\mu $ is called a principal null direction when
$
C_{\mu\nu\rho\sigma}k^\sigma = 0. 
$
Then 
$
L_Y \Big[ C_{\mu\nu\rho\sigma} k^\sigma \Big] = C_{\mu\nu\rho\sigma} L_Y k^{\sigma} = 0.
$  
Assuming that the spacetime is non-flat (which excludes, e.g. the Minkowski space)
the Weyl tensor is non trivial, allowing us to conclude that the Lie derivative of the null direction should be again a null direction. 
Now a pp wave is is known to be of Petrov type N and have just one principal null direction, namely our ``vertical vector'' $\xi$. Therefore $L_Y\xi$ is proportional to $\xi$ itself,
\beq
L_Y\xi=\alpha(X^\mu)\,\xi
\label{Hallcond}
\eeq
where
 $\alpha(X^\mu)$ is a function. One can only know it if the conformal vector and spacetime are given. This is as far as that we can go with no further assumptions.
The chrono-projective property (\ref{Chronocond}), i.e.,
$
\alpha(X)=\psi=\const
$
 may or may not be satisfied at this level, as it is manifest from (\ref{MaMaLie}). 
 
We just mention that calculating the components of the Weyl tensor could lead to an alternative proof of our statement.

%%%%%%%%%%%%%%%%%%%%%%%%%%%%%%%%%%%%%%%%%%%
\subsection{The ``Bargmann'' point of view}\label{BargSec}
%%%%%%%%%%%%%%%%%%%%%%%%%%%%%%%%%%%%%%%%%%%

Further insight can be gained using  the 
``Bargmann'' framework \cite{Bargmann,DGH91}.
We first recall that the space-time of a $4$-dimensional gravitational wave  with metric (\ref{Bmetric}) we denote by $(\cM, g_{\mu\nu})$  can be viewed as the ``Bargmann space'' for a non-relativistic system in $2+1$ dimensional non-relativistic spacetime, obtained by factoring out the integral curves of the covariantly constant ``vertical'' vector $\xi=\p_V$.  The $4$ dimensional  Bargmann manifold will be referred  to as ``upstairs'' and the underlying non-relativistic $2+1$-d system will be  ``downstairs''.

The relativistic (metric)  structure of Bargmann space projects to a non-relativistic Newton-Cartan structure \cite{Bargmann}. 
The factor space has coordinates $(U,\bX)$ with 
$U$ playing the r\^ole of non-relativistic time ; the classical motions ``downstairs" are the projections of the null geodesics ``upstairs". See  \cite{Bargmann,DGH91} for precise definitions and details.

For example, the null geodesics of 4D flat Minkowski spacetime
 written in light-cone coordinates project to free non relativistic motions in (2+1) dimensions.    

 More generally, let us consider\footnote{
 In 4D, the  metric of  any  solution of the vacuum Einstein equations $R_{\mu\nu}=0$ which is conformal to some vacuum Einstein solution can be brought to the form 
 \cite{Brinkmann,DGH91}, 
\beq
ds^2 = G_{ij}(U,\bX)dX^idX^j  + 2 dU dV - 2\Phi (U,\bX)\, dU^2\,. 
\label{Gen4DB} 
\eeq
where $G_{ij}(U,\bX)$ is a possibly $U$ [but not $V$] dependent metric on transverse space.
This is however  \emph{not} true in $D\geq5$ dimensions, 
allowing for more freedom \cite{Brinkmann,DGH91}.}, 
\beq
ds^2 = d \bX^2  + 2 dU dV - 2\Phi(U,\bX)\, dU^ 2\, 
\label{flattransv4D}
\eeq
The geodesics are described by the action 
$
S=\displaystyle\int \!\cL_{geo}\,d\sigma
$ 
with $\cL_{geo}$ in (\ref{geoLagrangian}).
The equations of motion are 
\beq
\ddot \bX =-(\dot U)^2 \frac{\partial \Phi}{\partial \bX}, 
\qquad 
\ddot U = 0, 
\qquad 
\frac{\;d}{d\sigma}\big(\dot{V} -2\Phi\, \dot{U}\big)  =  -\frac{\partial\Phi}{\partial U} \,\dot{U}^2\, .
\label{XUVeq}
\eeq
The related $4D$ geodesic Hamiltonian is
\beq
\cH= \frac{1}{2}{\bP}^2 +P_U P_V + \Phi(\bX, U) P_V^2\, ,
\label{Ham}
\eeq
where $\bm{P} = \dot{\bm{X}}$, $P_U = \dot{V} - 2\Phi\, \dot{U}$ and $P_V = \dot{U}$ is a constant.
The Hamiltonian (\ref{Ham}) and the Lagrangian
(\ref{geoLagrangian}) are in fact identical.
As they do not depend explicitly upon $\sigma$, we also  have the constraint 
\beq
{\dot \bX}^2 + 2\dot U \dot V - 2\Phi(\bX) (\dot U)^2 =-\epsilon
\label{constraint} 
\,,
\eeq 
where $\epsilon =1$ for timelike geodesics and $\epsilon=0 $ for null geodesics.

Focusing our attention at null-geodesics by requiring $\cH\equiv0$ and $P_V =M$
yields the non-relativistic $2+1-$d Hamiltonian ``downstairs'', 
\beq
 H_{NR} = \frac{\bm{P}^2}{2M} + M\Phi(\bm{X}, U) = - P_U\, .
\label{HNRPU} 
\eeq
$-P_U$ is the Hamiltonian for a non-relativistic particle of mass $M$ downstairs, and  $U=M\sigma$ plays the role of Newtonian time. $\Phi(\bX,U)$ is identified as a [possibly ``time''-dependent] scalar potential. The projected motion is governed by  the single equation in (\ref{XUVeq}).

As seen from (\ref{constraint}), the condition $\cH=0$ implies 
\beq
\dot{V} = -\dot{U} \Big( \frac{1}{2}\frac{\bm{\dot{X}}^2}{\dot{U}^2} - \Phi(X, U) \Big)
=- \left(\frac{1}{2M} \left(\frac{d\bX}{dU}\right)^2 -M\Phi(\bm{X}, U)\right) =- L_{NR}\,,
\label{dotVL}
\eeq
where $L_{NR}$ is the \emph{non-relativistic Lagrangian}. Therefore the  vertical coordinate is essentially (minus) \emph{the classical action} along the path $\bX(\sigma)$,
\beq
V=V_0-S,\quad S=\int\! L_{NR} \ d\sigma,
\label{Vevol}
\eeq
as noticed already by Eisenhart \cite{Eisenhart}.

When the potential $\Phi$ happens not to depend on $U$ explicitly, $\p\Phi/\p U=0$,
eqn. (\ref{XUVeq}) implies that $(\dot{V}-2\Phi\, \dot{U})$ is also conserved; eliminating $\dot{V}$ using (\ref{dotVL}) 
 yields the constant of the motion
$E=\half\left({\dot\bX}/{\dot U}\right)^2 + \Phi ,
$
 identified as the conserved  energy for unit mass of the projected motion. 
The special choice (\ref{Bmetric}),
 \beq
 \Phi(U,\bX)=-\half K_{ij}(U){X}^i{X}^j
 \eeq
 where $K_{ij}$ is a traceless symmetric matrix, represents, in Bargmann terms, a \emph{time-dependent anisotropic (attractive or repulsive) harmonic oscillator in the transverse plane}  \cite{Carroll4GW,OurMemory,POLPER}.

\goodbreak

%%%%%%%%%%%%%%%%%%%%%%%%%%%%%%%
%\section{Non-relativistic
%Bargmann conformal transformations} \label{NRconf}
%%%%%%%%%%%%%%%%%%%%%%%%%%%%%%%

For a general Bargmann space, those isometries 
(resp. conformal transformations)
 which preserve in addition the vertical vector $\xi=\p_V$, i.e.,  which satisfy (\ref{isom}) resp. (\ref{confom}),  with the additional condition
\beq
f_{*} \xi = \xi
\qquad\text{infinitesimally}\qquad
L_Y\xi=0
\label{xifix}
\eeq
 span the [generalized] \emph{Bargmann} (alias extended Galilei) 
resp. the [generalized] \emph{extended Schr\"odinger} \emph{group/algebra}. One can prove that the  conformal factors $\Omega$ resp. 
$\omega$ depend only on $U$ \cite{Bargmann,DGH91}. 
The homothety (\ref{homothety0})  belongs  to the [extended] \emph{chrono-projective} group \cite{5Chrono,DGH91,DuLaz} \footnote{The original definition \cite{DThese} is in the Newton-Cartan structure of non-relativistic spacetime.}
 defined, in general, by weakening the constraint (\ref{xifix}), \vskip-6mm
\besub
\begin{align}
&f^*g_{\mu\nu}= \Omega^2(U)g_{\mu\nu} 
\qquad\text{\small infinitesimally}\qquad
L_Yg_{\mu\nu}=2\omega(U)\,g_{\mu\nu}
\label{confombis}
\\
&f_{*} \xi\quad = \Psi\, \xi\quad\quad
\qquad\;\;\text{\small infinitesimally}\qquad
L_Y\xi\quad=\quad\psi\,\xi
\label{chronoxi}
\end{align}
\label{chronoprojdef}
\esub
where  $\Psi$ resp. $\psi$ are constants. It is a further 1-parameter (non-central) extension of the (centrally extended) Schr\"odinger group.

%%%%%%%%%%%%%%%%%%%%%%%%%%%%%%%%%%%%%%%%%%%%%%%%
\section{The Minkowski case}\label{MinkowskiSec}
%%%%%%%%%%%%%%%%%%%%%%%%%%%%%%%%%%%%%%%%%%%%%%%%

Here we list the generators of the conformal transformations for \emph{flat  Minkowski spacetime}  $d\bX^2+2dUdV$. 
 Plane gravitational waves with non-trivial profile $K_{ij}$ will be studied in sec. \ref{BJRSec}.
The flat metric can, alternatively, be thought as the Bargmann space for a $2+1$ dimensional NR particle (\ref{HNRPU}). The  various subalgebras/subgroups can be identified by listing the
 generators of $\Ort(4,2)$ \cite{HHAP}, 
\vskip-6mm
\begin{subequations}
\label{confalg}
\begin{align}
Y_U &= \partial_U, \qquad Y^i_T = -\partial^i, \qquad Y_V = -\partial_V, \quad\, \qquad \qquad \qquad\text{translations},  
\\
Y_{12} &= X^1\partial^2-X^2\partial^1, \qquad \qquad \qquad \qquad \qquad \qquad \qquad \qquad\text{$X^1-X^2$ rotation}, 
\\
Y^i_{B} &= U\partial^i - X^i\partial_V,  \qquad \qquad \qquad \qquad \quad \qquad \qquad \qquad \quad\;\;\text{galilean boosts}, \\
Y^i_{AB} &= X^i\partial_U - V\partial^i,  \qquad \qquad \qquad \qquad \quad \qquad \qquad \qquad \quad\;\text{``antiboosts''},
\label{antib} \\
Y_{UV} &= U\partial_U - V\partial_V,   \qquad \qquad \qquad \qquad \quad \qquad \qquad \qquad \quad\;\;\text{U-V boost}, 
\label{UVboostM}
\\
Y_{D} &= 2U\partial_U + X^i\partial^i,  \qquad \qquad \qquad \quad \quad \qquad \qquad \qquad \qquad\text{Sch dilatation}, 
\label{dilationM}\\
Y_{K} &= U^2\partial_U+UX^i\partial^i-\frac{\bm{X}^2}{2}\partial_V, \qquad \qquad \qquad \qquad \qquad \quad\;\text{Sch expansion},
\label{expansionM} \\
Y_{C1} &= \frac{\bm{X}^2}{2}\partial_U -VX^i\partial^i - V^2\partial_V, \qquad \qquad \qquad \qquad \qquad \quad\;\text{$C_1$},
\label{C1} \\
Y^i_{C2} &= X^i U\partial_U + X^i V\partial_V - \Big(\frac{\bm{X}^2}{2} +UV \Big)\partial^i + X^i (X^j\partial^j) \quad\;  \text{$C^i_2$}. 
\label{C2i}
\end{align}
\end{subequations}   

\begin{itemize}
\item
The 4D {\bf{Poincar\'e group}} $P_4$ is the $10-$parameter group of isometries generated by 
\beq
  \Big\{Y_U, \ Y^i_T, \ Y_V, Y_{12}, \ Y^i_{B}, \ Y^i_{AB}, Y_{UV} \Big\}\,,
\eeq
some of which do not preserve the vertical vector $\partial_V$,
$
[Y^i_{AB}, \partial_V] =\partial^i, \; [Y_{UV}, \partial_V] = \partial_V.
$

\item
The  isometries which \emph{do} preserve the vertical vector $\xi=\p_V$ (\ref{xifix}) provide us with the $7-$parameter  {\bf{Bargmann group}} \cite{Bargmann,DGH91} whose
Lie algebra defined by $({\cal{L}}_Y g)_{\mu\nu}= 0, \; [Y, \partial_V] = 0$ 
is spanned by 
\beq 
\Big\{Y_U, \ Y^i_T, \ Y_V, Y_{12}, \ Y^i_{B} \Big\}. 
\label{Bargalg}
\eeq

\item
The {\bf{extended Schr\"odinger} group} includes conformal transformations which preserve the vertical vector $\partial_V$.  The additional non-isometric transformations are \emph{non-relativistic dilations} $Y_{D}$ (\ref{dilationM}) and \emph{expansions} $Y_{K}$ (\ref{expansionM}) \cite{Schrgroup} which act as
\begin{subequations}
\begin{align}
({{\cal{L}}_{Y_{D}}} g)_{\mu\nu} &= 2\Lambda\ g_{\mu\nu} , \qquad [Y_{D}, \partial_V]=0, \\
 ({{\cal{L}}_{Y_{K}}} g)_{\mu\nu} &= 2\Lambda U \ g_{\mu\nu} , \quad\; [Y_{K}, \partial_V]=0. 
\end{align}
\label{DKextS}
\end{subequations}
The extended Schr\"odinger group has thus 9 parameters, namely,
\beq 
\Big\{Y_U, \ Y^i_T, \ Y_V, Y_{12}, \ Y^i_{B},\ Y_{D},\ Y_{K}  \Big\}. 
\label{extSalg}
\eeq
A detailed discussion  can be found in \cite{Bargmann,DGH91}.

\item The  {\bf{chrono-projective group}} \cite{DThese,5Chrono,DGH91,DuLaz,NCosmo} is a subgroup of the [relativistic] conformal group, (\ref{confom}), defined by the  condition 
$[Y, \partial_V] = \psi\, \partial_V$, cf. (\ref{chronoxi}). For $4D$ Minkowski space it has $10$-parameters and is spanned by  \cite{5Chrono,DGH91,DuLaz, NCosmo} 
\beq 
\Big\{Y_U, \ Y^i_T, \ Y_V, Y_{12}, \ Y^i_{B},\ Y_{D},\ Y_{K},\ Y_{UV} \Big\}. 
\label{chronoalg}
\eeq
The $U-V$ boost $Y_{UV}$ in (\ref{UVboostM}) is, in particular, a \emph{chrono-projective isometry}~: it satisfies (\ref{chronoprojdef})  
with $\omega=0$ and $\psi=1$, respectively.  
The homothety (\ref{infhomo}) can be expressed as 
\beq
Y_{hom}=Y_D-2Y_{UV},
\label{homoDuv}
\eeq
 therefore  belongs to the chrono-projective algebra. 
 
These  ``Bargmannian'' expressions $Y_K, Y_{UV}$ and $Y_{hom}$ appear in the literature independently, under the name of ``special conformal Killing vectors of the pp wave spacetime'' \cite{MaMa,KeaTu04}. 

\item The $(2+1)$D {\bf{Carroll group}} \cite{Leblond,DGH91,Carrollvs,NewCarroll} is the restriction of the Bargmann group to the  $3$D submanifold $\cC$ defined by the constraint $U=0$,
 \beq
f^*g = g,
\qquad
f_{*}\xi=\xi\,,
\qquad
f(\cC) \subset \cC\, .
\label{Carrolldef}
\eeq
It is a $6$ parameter subgroup embedded into $P_4$.  $U-$translations $Y_U$ are no more allowed.
 Its generators are, 
\beq
\Big\{Y^i_T,\ Y_V,\ Y^i_{B},\ Y_{12}\Big\}, \quad U=0.
\eeq

The Bargmann framework, whose primary aim is to provide a ``relativistic" description of non-relativistic physics,  has additional bonuses. One of them is to consider, instead of \emph{projecting} from 4 to 3 dimensions,  the
 \emph{pull-back} of a given Bargmann metric to the $3$-dimensional submanifold $U=0$ \footnote{The embedding  $U=\const$ would yield an equivalent construction. See also \cite{HallSteele}.}. 
${\cal C}$  has coordinates $(\bX,V)$ and carries  a \emph{Carroll structure}. $\xi=\p_V$ ; the coordinate $V$ is interpreted as \emph{``Carrollian time''} \cite{DGH91,Carrollvs}.

\item
The {\bf{Schr\"odinger-Carroll group}} is the conformal extension of the Carroll group \emph{within the conformal group} $\Ort(4,2)$, obtained by relaxing the isometry condition in (\ref{Carrolldef}) but still requiring that $\partial_V$ be preserved,
\beq
({\cal{L}}_Y g)_{\mu\nu}= \Omega^2 g_{\mu\nu}, \qquad 
[Y, \partial_V] = 0\,, \qquad U=0 \,.
\eeq
It has $8$ generators, namely those of the Carroll isometries, augmented by non-relativistic dilations and expansions,
\beq
\Big\{Y^i_T,\ Y_V,\ Y^i_{B},\ Y_{12},\ Y_{D},\ Y_{K} \Big\}, \quad U=0.
\eeq
and is represented by vectorfield  
\beq
 (\omega^i_j\,X^j+\gamma^i+\lambda{}\,X^i)\frac{\partial}{\partial X^i} + T(\bX) \p_V,\qquad
T(\bX) =\nu-\bbeta\cdot\bX
+\half\kappa\,\bX^2,
\label{Sch-carr}
\eeq
where  $T(\bX)$ is called, borrowing the Bondi-Metzner-Sachs (BMS) - inspired terminology \cite{BMS=confCarr,confCarroll},
\emph{supertranslations}. 

\item
The {\bf{Chrono-Carroll}} group is  a $1$-parameter extension of the Schr\"odinger-Carroll group with the weakened condition (\ref{chronoxi}), 
$ 
[Y,\partial_V] = \psi \,\partial_V.
$ 
This adds $Y_{UV}$ to the Schr\"odinger-Carroll algebra, yielding 9 generators 
\beq
\Big\{Y^i_T,\ Y_V,\ Y^i_{B},\ Y_{12},\ Y_{D},\ Y_{K},\ Y_{UV} \Big\}, \qquad U=0.
\eeq 
Infinitesimally,  (\ref{Sch-carr}) is  generalized to 
$
Y_V ={\psi}V +\nu-\bbeta~\cdot~\bX
+\half\kappa\,\bX^2\,.
$
\end{itemize}

\goodbreak

%%%%%%%%%%%%%%%%%%%%%%%%%%%%%%%%%%%%%%%%%%%%%
\section{Geodesics and their symmetries}\label{JacobiSec}
%%%%%%%%%%%%%%%%%%%%%%%%%%%%%%%%%%%%%%%%%%%%

In this section we revisit some aspects of geodesics 
and the conserved quantities associated with Killing and  reps. conformal Killing vectors.

%%%%%%%%%%%%%%%%%%%%%%%%%%%%%%%%%%%%%%%%%%%%
\subsection{Affinely parametrised geodesics}
%%%%%%%%%%%%%%%%%%%%%%%%%%%%%%%%%%%%%%%%%%%%

A fully covariant action for a
particle (and the only one for a massless particle) is
\beq
S = \int g_{\mu \nu} \frac{dX^\mu}{d \sigma}\frac{dX^\nu}{d \sigma}
d \sigma\,.
\eeq
Variation w.r.t.  $X^\mu$  gives the geodesic equations in the form
\beq
\frac{d^2 X^ \mu}{d \sigma ^2} + \Gamma^\mu _{\alpha \beta} \
\frac{d X^\alpha}{d \sigma} \frac{d X^\beta }{d \sigma} =0.
\label{sigmageo} 
\eeq
Here the $\Gamma^\mu_{\alpha \beta}$ are
the Christoffel symbols of the metric $g_{\mu \nu}$.
Because  there is no explicit dependence on $\sigma$,  we have the constraint (\ref{constraint}).
Choosing $\epsilon$ to be  $-m^2 \le 0$, 

$\bullet$ When $m\ne0$,
one sees that 
\beq
|m|d \sigma = d \tau,
\eeq
where $m$ is the relativistic mass and
 $\tau $ is proper time along the curve 
$X^\mu =X^\mu(\sigma)$.

$\bullet$  However when our geodesic is massless, $m^2 =0$, then $\sigma$ is called an \emph{affine parameter}, and is defined only up to an affine transformation.

The constraint (\ref{constraint}) may be written as 
$ 
g^{\mu \nu} P_\mu P_\nu = -m ^2\,,  
$ 
where
$P_\mu = g_{\mu \nu} \frac{dX^\nu}{d \sigma}$ is the 4-momentum.
If $m^2 \ne 0$ one has $P_\mu = |m| g_{\mu \nu} \frac{dX^\nu}{d \tau}$. 
In flat space one may set $P_0=-E$, and obtains the well known formulae
$ 
E^2-\bP^2 = m^2\,,  \,  E= \sqrt{m^2 + \bP^2}\,.
$ 

For the  general plane gravitational wave (\ref{Bmetric}) the constraint is
\beq
- K_{ij}(U) X^iX^j P_V^2 +2P_U P_V + m^2 + P_iP_i =0  \,.   
\label{PVgen}
\eeq
In general the only conserved quantity is $P_V$. If in addition $K_{ij}$ is independent of $U$, we have an additional conserved quantity, 
\beq
P_U = \frac{1}{2 P_V} \bigl (K_{ij}X^iX^j P_V^2 - m^2 - P_iP_i\bigr) \,.  
\eeq

We mention for completeness  that null geodesics lying in the null hypersurfaces $U=\const$, referred to as the \emph{null geodesic generators of the null hypersurfaces} $U=\const$ ; they may be related to lifts of isotropic geodesics in Newton-Cartan spacetimes \cite{DThese,confCarroll} for which
 $V$ is an affine parameter.

%%%%%%%%%%%%%%%%%%%%%%%%%%%%%%%%%%%%%%%%%%%%%%%%%%%
\subsection{Killing resp. conformal Killing vectors}
%%%%%%%%%%%%%%%%%%%%%%%%%%%%%%%%%%%%%%%%%%%%%%%%%%%

We first recall what happens for \emph{Killing vectors}.
If we define the tangent vector of a curve with general
parameter $\lambda$ by  $T^\mu = \frac{dX^\mu}{d \lambda}$,
then a geodesic satisfies
\beq
T^\alpha T^\mu _{\; ;\, \alpha} = h(\lambda)\, T^\mu
\label{lambdageobis}
\eeq
for some function $h(\lambda)$, where the
``\,semicolon $; \alpha $'' denotes covariant derivative.  
\goodbreak

\kikezd{Killing vectors}.

Suppose first that $Y^\mu $ is a \emph{Killing vector field}; then
it satisfies Killing's equations
\beq
Y_{\mu ; \alpha} + Y_{\alpha ; \mu}= 0 \,.
\label{Killingeq}
\eeq\vskip-5mm
It follows that
\beq
{\cal E}=  Y_\mu T^\mu = g_{\mu \nu}T^\mu Y^\nu \,
\qquad\text{satisfies}\qquad
{\cal E}_{; \alpha}T^\alpha =
\frac{d {\cal E}}{d \lambda} = h(\lambda)\, {\cal E}\,.
\label{KillCons}
\eeq
Then we get a conserved quantity for the geodesic motion,
\beq\bigbox{
\cQ_Y=
g_ {\mu \nu}\frac{dX^\mu}{d \sigma} Y^\nu   =  g_{\mu \nu} \frac{dX^\mu}{d \lambda} \frac{d \lambda}{d \sigma}\, Y^\nu\,,\quad
\frac{d\cQ_Y}{d\sigma}=0\,.
\;}
\label{KillingCons}
\eeq
Translations along the ``vertical'' vector  $\xi=\p_V$ are isometries for any metric of the form (\ref{Bmetric}). 
The associated conserved quantity  $P_V= M$  in (\ref{PVgen})
is identified, in the Bargmann framework, with the mass downstairs (as mentioned in section \ref{BargSec}).

\goodbreak
%%%%%%%%%%%%%%%%%%%%%%%%%%%%%%%%%%%
\kikezd{Conformal Killing vectors}.
%%%%%%%%%%%%%%%%%%%%%%%%%%%%%%%%%%%

Now we suppose that we have instead a \emph{conformal Killing vector} $Y^\mu$, i.e., one for which 
\beq
Y_{\mu ;\nu}+ Y_{\nu ; \mu}= 2\omega\, g_{\mu \nu}  
\eeq 
for some function $\omega$. If $\omega={\rm constant}$, $Y^\mu$ is called a \emph{homothetic Killing vector} since it generates a homothety.
For a \emph{timelike geodesic} with tangent vector  
$ 
T^\alpha = \frac{dx^\alpha}{d \tau} \,
$ 
where $\tau$ is proper time along the geodesic
so that
$ 
T^\alpha T_{\alpha} =-1\,,
$ 
we have instead
\beq 
(Y_\alpha T^\alpha)_{\,;\, \mu }T^\mu =-\omega\,.
\eeq
Thus in general the quantity  (\ref{KillCons}) i.e.
$
Y_\alpha T^\alpha= \frac{1}{m}Y^\mu P_\mu
$ 
(where $P_\mu=m T_\mu$ is the momentum of a particle of mass $m$) is \emph{not} constant
along the world line. From the point of view of the covariant
Hamiltonian treatment, $Y^\mu P_\mu$ is the moment map generating the lift to the
co-tangent bundle of the conformal transformation  of the base manifold\,.

In the special case of a homothety when $\omega =\omega_0=\const$, we find that
$   
\frac{d (Y_\alpha T^\alpha)}{d \tau} = -\omega_0 
\,\Rightarrow\,
Y_\alpha T^\alpha = -\omega_0 \tau  -\omega_{-1}
\,.
$  
Alternatively, deriving again, we have
$ 
{d^2 (Y_\alpha\, T^\alpha)}/{d\tau^2}  = 0\,,
$ 
which is a covariant version of the \emph{Lagrange-Jacobi identity} \cite{JacobiLec}.

Conformal Killing vectors do not generate symmetries for timelike geodesics. However, as observed by Jacobi \cite{JacobiLec}, while $Y_\alpha T^\alpha$ is not in general conserved, the two constants of integration above yield, in modern language, the conserved quantities (\ref{ConsSch}) associated 
with non-relativistic dilations and expansions, respectively  \cite{DHNC}.

%%%%%%%%%%%%%%%%%%%%%%%%%%%%%%%%%%%%%%%%%%%%%%%%%%%
\subsection{Conserved quantities for null geodesics}\label{nullCQ}
%%%%%%%%%%%%%%%%%%%%%%%%%%%%%%%%%%%%%%%%%%%%%%%%%%%

By contrast, if one considers an affinely
parametrised \emph{null geodesic} with tangent vector
$ 
l^\alpha = {dx^\alpha}/{d\sigma}
$ 
that satisfies 
\beq
g_{\mu \nu} l^\mu l^\nu =0 \,,\qquad
 l^\mu_{\; ;\,\nu}\,l^\nu=0
\,,  
\eeq
we \emph{do obtain a constant of the motion},
\beq\medbox{
\cQ_Y= Y_\mu\, l^\mu \,,\qquad
\frac{d \cQ_Y}{d \sigma}= 0\,. \;
}
\label{nullQ}  
\eeq
\goodbreak

\vskip2mm
$\bullet$ As a first illustration, we re-derive the conserved quantities associated with Schr\"odinger dilations and expansions. For  $\Phi(\vX)=0$ and $\Phi(\vX) \propto |\bX|^{-2}$ (\ref{flattransv4D}) describes a free particle and the inverse-square potential respectively. The  generators $Y_D$ and $Y_K$ (\ref{DKextS}) are conformal Killing vectors.  Following the procedure outlined in sect. \ref{BargSec}, \eqref{nullQ} yields the conserved Schr\"odinger quantities downstairs \cite{Schrgroup},
\besub
\begin{align}
&\cD=P_i X^i-2E U &\text{dilation}
\label{Consdilat}
\\[4pt]
&\cK=-EU^2+UP_iX^i-\frac{M}{2}X_i X^i
&\text{expansion}
\label{Consexp}
\end{align}
\label{ConsSch}
\esub
These quantities are conserved for null geodesics ``upstairs'' and  project   to  well-defined conserved quantities for the projected non-relativistic motion. In fact $\cD, \cK$ close, with the projected Hamiltonian $H_{NR}$ to an ${\rm o}(2,1)$ algebra \cite{Schrgroup}. 
 
More generally, for motion along null geodesics eqn. (\ref{nullQ}) associates a conserved quantity to each conformal vector $Y$;   if the latter preserves in addition also the ``vertical'' vector $\xi=\p_V$,
$L_Y\xi=0$
(\ref{xifix}),  this quantity (we call of the Schr\"odinger type)  projects to a conserved quantity for the underlying non-relativistic dynamics ``downstairs" --- this is in fact the original idea of the Bargmann framework  \cite{Bargmann,DGH91}.
\goodbreak

%%%%%%%%%%%%%%%%%%%%%%%%%%%%%%%%%%%%%%%%%%%%%%%%%%
\section{Scalings as chrono-projective transformations} \label{homoSec}
%%%%%%%%%%%%%%%%%%%%%%%%%%%%%%%%%%%%%%%%%%%%%%%%%%%

Now we present a systematic and detailed discussion of scale transformations in the Bargmann framework.
  We start with the homothety (\ref{homothety0}) -- (\ref{infhomo}). Being a  conformal vector for  the gravitational wave spacetime, (\ref{nullQ}) provides us with 
\beq
\cQ_{hom} = X^i P_i+2V P_V\,,
\label{Qhomot}
\eeq
where $P_V$ is associated with the ``vertical'' Killing vector $\p_V$. $\cQ_{hom}$ is
conserved  for \emph{null} (but not for timelike) geodesics, as confirmed also by using the equations of motion (\ref{ABVeq}).

Assuming that the transverse motion $X^i(\sigma)$ had already been determined, the conservation of $\cQ_{hom}$ allows us to determine the evolution of the ``vertical coordinate'' ,
\beq
V(\sigma)=\frac{\cQ_{hom}}{2P_V}-\frac{X^i(\sigma)P_i(\sigma)}{2P_V}=
\frac{\cQ_{hom}}{2P_V}-\frac{1}{4P_V}\frac{\,d}{d\sigma} \big(X^i(\sigma)X_i(\sigma)\big)\,.
\label{VUGW}
\eeq

As explained in sec.\ref{BargSec},  the null dynamics in 4D projects, in the Bargmann framework, to an underlying non-relativistic system in $2+1$D, whereas $P_V$ becomes the mass, $M$ ; $U$ becomes the non-relativistic time. The non-relativistic Hamiltonian and Lagrangian are recovered as in (\ref{HNRPU}) and (\ref{dotVL}), allowing us to express   
\beq
\cQ_{hom}=Q_{NR}+2MV_0,\qquad
Q_{NR} = X^i P_i-{2} \int^U\!\! L_{NR} (u) d u\,.
\label{QNR}
\eeq 
$Q_{NR}$ is thus  \emph{conserved ``downstairs'} (as it can be confirmed  directly using the eqns of motion). 

More generally, the anisotropic rescaling
\beq
\label{abcscale}
U \to \mu^b\,U,
\qquad
{X^i} \to  \mu^a\, X^i,
\qquad
V \to \mu^{c}\, V\,,
\qquad
\mu=\const
\eeq
induces, for the Brinkmann metric (\ref{Bmetric}),
$$
g_{\mu\nu}dX^\mu dX^\nu\to\mu^{2a}
\Big(\delta_{ij} dX^i dX^j + \mu^{-2a+b+c}\, 2 dU dV + K_{ij}(\mu^bU) \mu^{2b}X^i X^j dU^2\Big)\,.
$$
This is conformal provided
$ 
c=2a-b
$ and $K_{ij}(\mu^bU)=\mu^{-2b}K_{ij}(U) \,.
$ 
Then, for any $b$,
\beq
g_{\mu\nu}dX^\mu dX^\nu\to\Omega^2g_{\mu\nu}dX^\mu dX^\nu,
\qquad
\Omega=\mu^{a}.
\eeq
${\cal{L}}_Y \partial_V = -c\partial_V$ implies 
 that the vector field $Y$  is genuinly chrono-projective whenever $c=2a-b\neq 0$.
The associated conserved quantity 
\beq
\cQ_{a,b}=aX^iP_i+b\,UP_U+cVP_V
\label{abcharge}
\eeq
induces a conserved charge downstairs, 
\beq
Q_{a,b}=\cQ_{a,b}+cV_0P_V
=aX^iP_i-b\,U E-c\left(\int^U\!\!L_{NR}\right)M\,.
\label{Qdown}
\eeq
Note here the new ``chrono-projective'' term proportional to the non-relativistic action.

\benu
\item
 For $a=b=c=1$ we would get the  relativistic (isotropic) dilation $U\to \mu U,\, X^i \to \mu X^i,  V \to \mu V $; when $b=2a$ we get Schr\"odinger dilations.

\item
When $b=0$ we recover, for any profile $K_{ij}(U)$,  (\ref{QNR}).

\item
If $K_{ij}$ is $U$-independent (as for Brdi\v{c}ka metric    
 (\ref{Brdmetric}) below), then  $b=0$ ;
 
\item
$b\neq0$ could be obtained for the [singular] non-trivial profile \cite{Sippel, Eardley,MaMa, Keane}. (See also class 11 in Table 4 of \cite{KeaTu04}). 
\beq
K_{ij}(U)=\frac{K_{ij}^0}{U^2}\,,
\quad
K_{ij}^0=\const
\label{inverseU2}
\eeq
Choosing $a=0$  we obtain a \emph{chrono-projective  isometry}  
  -- namely our \emph{U-V boost} $Y_{UV}$,  $U\to \mu U,\, X^i \to X^i,  V \to \mu^{-1}V $. Its conserved charge is ``chrono-projective''
\beq
\cQ_{UV}=UP_U-VP_V=- UE+\int^U\!\!L_{NR}-V_0P_V.
\eeq  
Choosing instead $b=0$, we recover (as said above) the homothety~(\ref{QNR})~\footnote{ The  profile
(\ref{inverseU2}) is  symmetric also w.r.t. Schr\"odinger dilations.
This is not a surprise, though, because the latter is  a combination of an UV boosts and of homothety, as seen before.}.
Thus this example has again a maximal i.e. a 7-parameter chrono-projective algebra.
\eenu 
 
The quotient $z=b/a$ is also called the \emph{dynamical exponent} \cite{ConfGal,Henkel}. The typical relativistic value is $z=1$ ; for Schr\"odinger-type expressions $z=2$. The ``chrono-projective'' contribution to the associated conserved quantity $\cQ$ with an additional action integral term arises when  $z\neq2$ \cite{DHNC,ConfGal,Henkel,ZhHhydro}.

%%%%%%%%%%%%%%%%%%%%%%%%%%%%%%%%%%%%%%%%%%%%%%%%%%%%%%%%%%%%%%
\section{Chrono-projective transformations in BJR coordinates}\label{BJRSec}
%%%%%%%%%%%%%%%%%%%%%%%%%%%%%%%%%%%%%%%%%%%%%%%%%%%%%%%%%%%%%%

Having reviewed the Bargmannian aspects, now we turn to a systematic study of chrono-projective transformations  of the  gravitational wave metric \eqref{Bmetric} with a non-trivial profile $K_{ij}(U)$. 
The conformal transformations of  pp-waves have been
determined some time ago \cite{Sippel, Eardley, MaMa, Keane}. Below  we study them in our case of interest in a novel way.
Motivated by their utility to identify the isometries \cite{Sou73,Carroll4GW,OurMemory}, we switch to Baldwin-Jeffery-Rosen (BJR) coordinates $(u,\bx,v)$ \cite{BaJe},
\beqa
U = u \, , \qquad  
\bX = P(u)\,\bx \, ,\qquad V = v - \frac{1}{4}\bx \cdot \dot{a}(u) \bx \, , 
\label{BtoBJR}
\eeqa 
where $a(u) = P^\dagger (u) P(u)$, the $2\times2$ matrix $P=(P_{ij})$
being  a solution of the Sturm-Liouville problem \cite{%Gibb75,
OurMemory,SLC} 
\beq 
\ddot{P} = K P \, , \qquad P^\dagger \dot{P} = \dot{P}^\dagger P \,.  
\label{SLP}
\eeq 
Then  the metric \eqref{Bmetric} takes the form,
\beq 
g = a_{ij}(u)dx^idx^j + 2dudv\,. 
\label{BJRmetric}
\eeq
The conditions (\ref{chronoprojdef})  determine the form of the general chrono-projective vector field
\beq
\medbox{
Y = Y^u (x,u) \pa_u + Y^i(x,u) \pa_{i} + \big(b(x,u) - \psi\, v\big)\, \pa_v \,. 
}
\label{ChronoBJR}
\eeq
\goodbreak

\vskip2mm
 The conformal Killing equation  (\ref{chronoprojdef}) requires, \vspace{-3mm}
\besub
\begin{align}
&\partial_iY^u = 0\,,
\label{Yia}
\\
&\partial_u Y^u = 2\omega+\psi\,,
\label{Yib}
\\
&\p_uY^v = 0 
\label{Yic} 
\\
&\partial_i Y^v +(\partial_u Y^j)a_{ij}=0\,, 
\label{Yid}
\\
&Y^u (\partial_u a_{ij}) + a_{kj}(u)\partial_i \big(Y^k(x,u)\big)+ a_{ki}\big(\partial_j Y^k(x,u)\big) = 2\omega(u) a_{ij} \,. 
\label{Yie} 
\end{align}
\label{Yi}
\esub
Eqn. (\ref{Yia})  implies that $Y^u = Y^u(u)$ hence $\omega = \omega(u)$; then
(\ref{Yib}) can be solved as
\beq
Y^u(u)= {\epsilon}+ \int_0^u\!\big(2\omega(w)+\psi\big)dw\,.
\label{Yuompsi}
\eeq
($\epsilon=\const$)
Eqn. 
(\ref{Yic}) implies that 
 $b$ is $u$-independent, $b = b(x)$.
 Then  $Y^i(x,u)$ can be written as
$ 
Y^i(x, u) = K^i(u) + F^i(x) + L^i(x,u).
$  
 Substituting into  (\ref{Yid}) we get,
\beq
\label{csYigenex}
Y^k(x,u) = F^k(x) - H^{ki}(u)\partial_i b(x)\,,
\eeq
where $H^{ki}$(u) is Souriau's $2\times2$ matrix \cite{Sou73,Carroll4GW}, 
\beq
H^{ki}(u) = \int_{0}^u a^{ki}(w) dw \,,
\label{Smatrix}
\eeq 
where $(a^{ij})$ is the inverse matrix, $a^{ij}a_{jk}=\delta^i_k$\,.
Inserting this into (\ref{Yie}), the last condition can be written as
\begin{eqnarray}
\label{csmaineq}
&&-2\omega(u)  a_{ij}(u)+Y^u(u) (\partial_u a_{ij}(u)) + 
\\[6pt]
&&a_{kj}(u)\Big(\partial_i F^k(x) - H^{km}(u)  \partial_i \partial_m b(x)\Big) 
+ a_{ki}(u)\Big(\partial_j F^k(x) - H^{km}(u)\partial_j \partial_m b(x) \Big)=0.\nn
\end{eqnarray}

Collecting our results, 
%%%%%
\begin{subequations}
\begin{align}
&Y^u (u)  = \epsilon+ 2 \int^u\!\omega(w) dw + \psi u \,,
\label{BJRu}
\\
&Y^i (x, u)  = F^i(x) - H^{ij}(u)\partial_j b(x)\,,
\label{BJRi} 
\\
&Y^v (x, v)  = b(x) -\psi v\, .
\label{BJRv}
\end{align}
\label{csKvs}
\end{subequations}
Thus ${\epsilon}=\const$ is a time translation ; the conformal resp. chrono-projective factors $\omega$ and $\psi$ contribute to time dilations. 
Although the functions 
$F^i(x)$ and $b(x)$ are generally profile-dependent and can only be determined from (\ref{csmaineq}), we can conclude that
$b(x)$ is at most quadratic in $x$,
\beq
b(x)= b_{ij}x^ix^j-b_ix^i+h,\, \quad  b_i,\, h = \const.
\label{bBJR}
\eeq
Thus $F^i(x)$ should be at most of the first order in $x^j$, see our examples in the next section.

A particular transformation in (\ref{csKvs}) is $\psi(u\partial_u - v\partial_v)$, associated with the chrono-projective factor, we called before a  $u-v$ boost, cf. (\ref{UVboostM}).

%%%%%%%%%%%%%%%%%%%%%%%%%%%%%% 
\kikezd{Hamiltonian structure} 
%%%%%%%%%%%%%%%%%%%%%%%%%%%%%
 The geodesic  Lagrangian resp. Hamiltonian are,
 in BJR coordinates, 
\beq
\cL =  \frac{1}{2} a_{ij} (u) \dot{x}^i \dot{x}^j + \dot{u} \dot{v}\,,
\qquad
\cH = \frac{1}{2} a^{ij}p_i p_j + p_u p_v\,,
\label{BJRLagHam}
\eeq
where the canonical momenta $p_\mu=\p\cL/\p \dot{x}^{\mu}$ are
$ 
p_u = \dot{v}, \, p_v = \dot{u}, \, p_i = a_{ij} \dot{x}^j \,\Rightarrow \, \dot{x}^i = a^{ij}p_j\,.
$ 
By (\ref{nullQ}) the conserved quantity associated with the conformal vectorfield $Y$ is,
\beq
\medbox{
\cQ_Y = Y^\mu p_\mu = Y^u (x,u)\, p_u +Y^i(x,u)\, p_{i}  +  \big(b(x) - \psi v\big)\, p_v \,.}
\label{ChronoProjQ}
\eeq
By using the Poisson bracket,
$
\big\{\cR,\cT \big\} =\frac{\p\cR}{\p{x}^\mu}\frac{\p\cT}{\p{p}_\mu}-
\frac{\p\cR}{\p{p}_\mu}\frac{\p\cT}{\p{x}^\mu}\,,
$
the generating vector field is recovered as 
$ 
Y^{\mu}\p_\mu=\{x^{\mu}\, , \cQ_Y\}\p_\mu \,.
$ 
 Rewriting the Hamiltonian as,
\beq
\cH = p_\mu \dot{x}^\mu - \cL= \frac{1}{2} (a^{-1})^{ij} p_i p_j +p_u p_v = \frac{1}{2} (g^{-1})^{\mu\nu} p_\mu p_\nu\,,
\label{cHbis}
\eeq
we have
\beq
\big\{\cQ, \mathcal{H}\big \} = - \frac{1}{2} \mathcal{L}_Y{g}^{\mu\nu} p_\mu p_\nu,
\aand
\{\cQ, \xi\big\} = -\mathcal{L}_Y(\xi)\,.
\label{QHxiPB}
\eeq
By (\ref{chronoprojdef})
a charge  which is conserved along null geodesics should satisfy 
\beq 
\left\{\cQ, \cH \right\} = 2\omega\, \mathcal{H}\,,
\qquad
\big\{\cQ, \xi\big\} = -\psi\, \xi \, 
\eeq 
with $\omega=\omega(u),\, \psi=\const$.
These formulas come handy to check whether a given quantity is conserved or not.

\vskip-5mm
\goodbreak

%%%%%%%%%%%%%%%%%%%
\kikezd{Isometries and homothety}:
%%%%%%%%%%%%%%%%%%%
As  seen from (\ref{csmaineq}), one can not have a generic profile independent symmetry unless $Y^u = 0$. 
Setting $Y^u = 0$ correlates the conformal and chrono factors, $\omega = -\psi/2 = \const$ and $\epsilon = 0$. Then (\ref{csmaineq}) simplifies as
\beq
\partial_i F^k(x) - H^{km}(u)  \partial_i \partial_m h(x) = \omega \delta^k_i.
\eeq
Thus, we conclude that $b(x)=-b_i x^i + h$ and $F^i(x) = \omega x^i+f^i$. Putting all into (\ref{csKvs}), we obtain the combination $Y = Y_{iso} + Y_{hom}$
\begin{eqnarray}
\label{BJRiso}
Y_{iso} &=& f^i\partial_i + h\partial_v + b_i (H^{ij}\partial_j - x^i\partial_v) , \quad f^i = \text{const.},  \\
Y_{hom}&=& \omega (x^i\,\p_i+ 2v\,\p_v).
\label{BJRhom}
\end{eqnarray} 
$Y_{iso}$ contains the 5 standard isometries (namely $\bm{x}$-translations, $v$-translation and $\bx$-boosts, respectively), identified as the Carroll group with broken rotations \cite{Leblond,Carroll4GW,Carrollvs}, see sec.\ref{MinkowskiSec}.  
In the Hamiltonian framework, they become
\beq
T^i = \delta^{ij}\,p_j ,\quad  
T_v  =  p_v\,, \quad
B^i = H^{ij}p_j - x^i\, p_v \, , 
\label{5iso}
\eeq
and they all commute with the geodesic Hamiltonian $\cH$ (\ref{cHbis}) ; the only  non-vanishing brackets are \footnote{Note that this is not a central extension; the generators belong themselves to the algebra.}
\beq
\left\{T_i \,, \,B^j \right\} = \delta_i^j\, p_v \,.
\eeq
Being proportional to $\omega$,  $Y_{hom}$ is the homothetic vector field (\ref{infhomo})  exported to BJR coordinates ; it induces the 
conserved charge for null geodesics
\beq 
\cQ_{hom} = x^i p_i+2 v p_v  \, ,
\qquad  
\{Q_{hom} , \mathcal{H} \} = 2 \mathcal{H} \, . 
\label{homoHPB}
\eeq 

We would like to emphasize that the isometries (\ref{BJRiso}) and the homothety (\ref{BJRhom}) are valid for every profile $a_{ij}(u)$ and do not require any integrability equation. For comparison, we note that the homothetic Killing vector found in \cite{MaMa}, their eqn. $\#$ (48), is subject to an integrability condition.

In addition to the isometries and the homothety other (conformal) symmetries may arise; depending on eqn. (\ref{csmaineq}) (see the next section).

We just mention that the Ricci flatness of the Brinkmann metric  (\ref{Bmetric}), 
$\Tr(K_{ij})=0$, can be exported to BJR coordinates as \cite{BaJe, Sou73}
\beq
\Tr\left(\dot{L} + \frac{1}{2} L^2\right), \quad L = a^{-1}\dot{a}.
\eeq
So far we have not used this condition, therefore our solutions  (\ref{csKvs}) should apply for the special conformal Killing vectors of any pp-wave.

%%%%%%%%%%%%%%%%%%%%%%%%%%%%%%%%%%%%
\section{Examples}\label{Examples}
%%%%%%%%%%%%%%%%%%%%%%%%%%%%%%%%%%%%

Now we illustrate our  general theory on selected examples. 

%%%%%%%%%%%%%%%%%%%%%%%%%%%%%%%%%%%%
\subsection{Minkowski case}
%%%%%%%%%%%%%%%%%%%%%%%%%%%%%%%%%%%%

In the flat (Minkowski) case $a_{ij} = \delta_{ij}$ and $H^{ij}(u) = u\,\delta^{ij}$ the constraint (\ref{csmaineq}) requires
\beq
\label{csflatmaineq1}
-2u\partial_i\partial_j b(x) + \big(\partial_i F^j (x) + \partial_j F^i(x)\big) = 2\omega(u)\delta_{ij}
\eeq
and a simple calculation yields 
 the BJR form of the free chrono-projective Lie algebra  (\ref{chronoalg}),
\vspace{-7mm} 
\begin{subequations}
\begin{align}
&Y^u (u)  = \epsilon+ 2 \lambda\, u  +\kappa\, u^2 + \psi\, u 
\\
&Y^i (x, u)  = \omega^{i}_{\,j}x^j + f^i + ub_i +\lambda x^i +\kappa ux^i\,, 
\\
&Y^v (x, v)  = -\frac{\kappa}{2} \bx^2-\bb\cdot\bx + h -\psi v,
\end{align}
\label{freeCPalg}
\end{subequations}
where $\epsilon, \lambda,\kappa, \psi, \omega^{i}_{\,j}, f^i, b_i, h$ are constants.
Thus
 $\epsilon$ generates non-relativistic time translations, $\lambda$ and $\kappa$ are the parameters of Schr\"odinger dilations and expansions, $f^i$ and $h$ space and vertical translations, $\omega^{i}_{\,j}$  rotations, $b_i$ as Galilei boosts. 
$\psi$ generates  u-v boosts.

 In conclusion and consistently with what we said in sec. \ref{MinkowskiSec} we have  7 dimensional isometry group which satisfies
$
L_Yg_{\mu\nu}=0
$ and
$
L_Y\xi=0
$
and is identified with the Bargmann group (\ref{Bargalg}). The latter is extended to the [extended] Schr\" odinger
group (\ref{extSalg}) by the addition of Schr\"odinger dilations and expansions, which are conformal and leave the vertical vector invariant,
$$
L_Yg_{\mu\nu}=2\omega\,g_{\mu\nu}
\with
\omega=\lambda + \kappa u\,
\aand
L_Y\xi=0
$$
The 10-parameter chrono-projective group is then obtained by adding one more 
 transformation, for example u-v boosts. 
Alternatively, by (\ref{homoDuv}) u-v boosts can be replaced by the homothety.
 
As mentioned before, conformal Killing vectors of a plane gravitational wave, eqn. $\#$ (56) in \cite{MaMa}, are chrono-projective. However, flat pp-waves were excluded \footnote{The solutions (\ref{freeCPalg}) can be obtained from (\ref{MaMaW}) by putting  $H(u, x^A)=0$ in  eqn. $\#$ (55) of \cite{MaMa}. }, explaining why the full $o(4,2)$ algebra (\ref{confalg}) is not recovered~: ``antiboosts" (\ref{antib}) are, for example, not chrono-projective transformations.

Now we turn to examples with non-trivial profile.

%%%%%%%%%%%%%%%%%%%%%%%%%%%%%%%%%%%%%%%%%%%%%%%%%%%
\subsection{The Brdi\v{c}ka~ metric}\label{BrdickaSec}
%%%%%%%%%%%%%%%%%%%%%%%%%%%%%%%%%%%%%%%%%%%%%%%%%%%
 
Let us first consider the linearly polarized gravitational wave metric given in Brinkmann coordinates  \footnote{See also class 13 in Table 4 of \cite{KeaTu04}.} \cite{Brdicka},
\beq
dX_1^2+dX_2^2+2dUdV-2\Phi dU^2\,,
\qquad
\Phi=\half\Omega^2\Big(X_1^2-X_2^2\Big),\qquad
\Omega=\const\,
\label{Brdmetric}
\eeq
The potential $\Phi(\vX)$ is attractive in the $X_1$ and repulsive in the $X_2$ sector.

We note that
the Brdi\v{c}ka profile  is  $U$-independent, leaving us with the homothety as the only conformal symmetry.

We switch to BJR coordinates using the solution of the Sturm-Liouville equation (\ref{SLP})
\beq
P(u)=\diag\Big(\cos\Omega u\,,\, \cosh\Omega u\Big)\,.
\label{BrdminP}
\eeq
The induced Brinkmann $\to$ BJR transformation (\ref{BtoBJR}) i.e.
\footnote{Indices are lifted by the transverse metric, $x^i=a^{ij}x_i$.} 
\beq\left\{
\barraynb{llllll}
U&=&u \qquad  
X_1 = x^1 \cos(\Omega u)\qquad
X_2 = x^2\cosh(\Omega u)
\\[6pt]
V &=& v+\dfrac {\Omega}{4}(x^1)^2 \sin(2\Omega u)-\dfrac {\Omega}{4} (x^2)^2 \sinh(2\Omega u)
\earraynb\right.
\label{brdickatraf}
\eeq
yields the BJR metric (\ref{BJRmetric}) resp. Souriau matrix
\besub
\begin{align}
a(u)&=
 \diag\Big(\cos^2 (\Omega u)\,,\, \cosh^2 (\Omega u) \Big)\,,
 \label{BrdBJminRprof}
 \\
\big(H^{ij}(u)\big) &=
{\Omega}^{-1} 
\diag \big(
\tan(\Omega u) \,,\,\tanh(\Omega u)
\big).
\label{BrdSmat}
\end{align}
\esub

\begin{figure}[h]
\includegraphics[scale=.145]{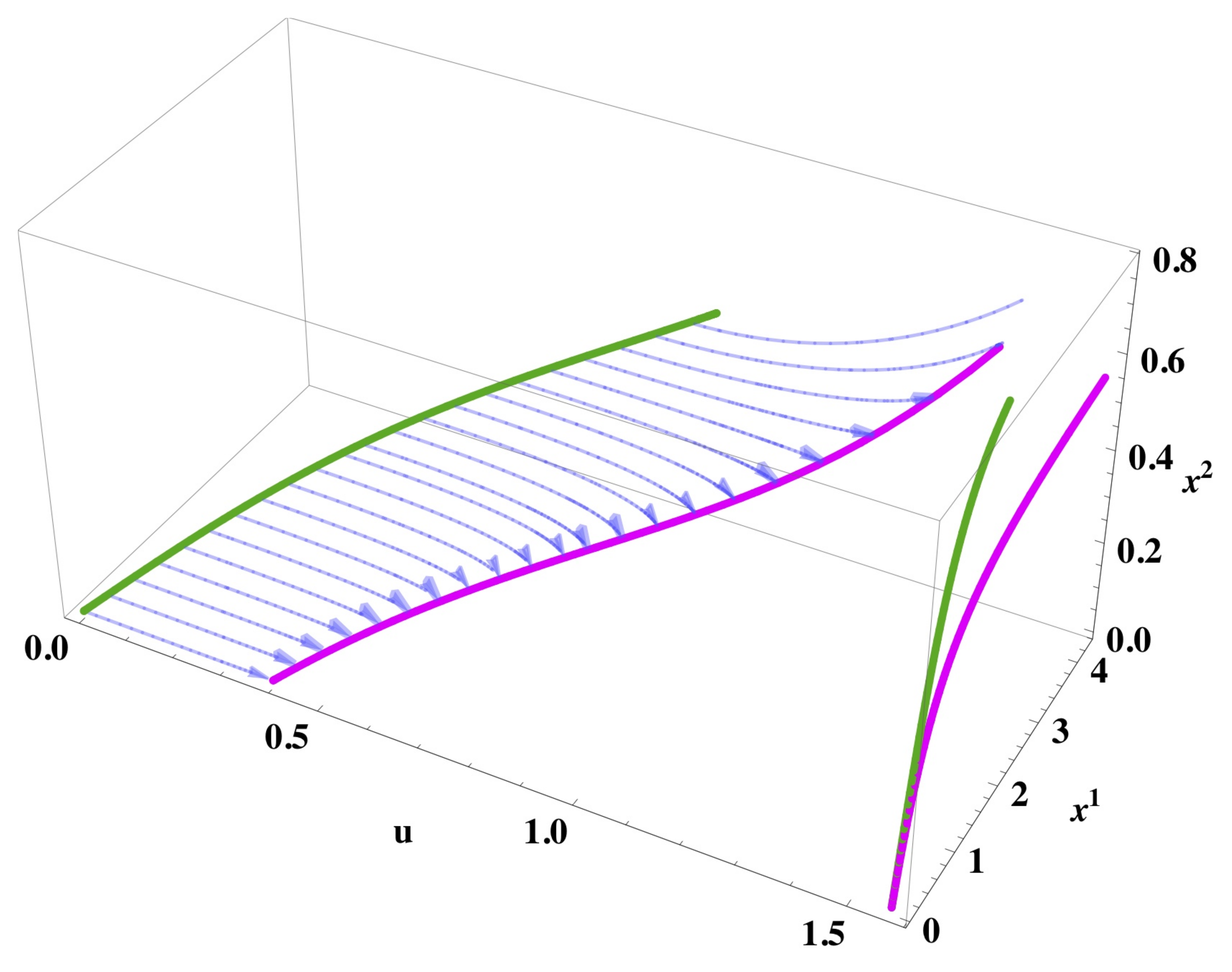}
\vskip-4mm
\caption{\textit{\small 
The $U$-translation in Brinkmann coordinates (\ref{Utr})
becomes, in BJR coordinates, ``screwed''.
}}  
\label{Brdiscrew}
\end{figure}

\vskip-5mm
\kikezd{A ``screw-type'' isometry}\footnote{We borrowed the word from
\cite{POLPER}, where a broken $U$ translation combines with a broken rotation into a symmetry, which acts as a ``screw'' \cite{exactsol,Carroll4GW,Ilderton}, as  illustrated in fig.\ref{CPPscrewfig} below.}.
When written in Brinkmann coordinates, the metric (\ref{Brdmetric}) is $U$-independent, implying that the $U$-translations 
\beq
U\to U + e
\label{Utr}
\eeq 
add a 6th manifest isometry to the $5$ standard ones. It is redundant nonetheless instructive to see how this comes about  in BJR coordinates~: $u\equiv U$-translations are still symmetries  but the implementation becomes distorted,  see fig. \ref{Brdiscrew}.

%%%%%%%%%%%%%% 
\goodbreak

To find all chrono-projective vectorfields we follow the recipe outlined in sec.\ref{BJRSec}~: starting with the general equations (\ref{csKvs}),
 a calculation similar to the one in the free case shows that
\beq\left\{\barraynb{lll}
\ddot{\omega} + 2 \Omega^2 (2\omega + \psi) &=& 0
\\
\ddot{\omega} - 2 \Omega^2 (2\omega + \psi) &=& 0 
\earraynb\right.
\label{homoscales}
\eeq
whose consistency requires $2\omega+\psi = 0$. 
Therefore $\omega$ is a constant and $Y^u$ is a mere $u$-translation,  $Y^u=\epsilon$.
Then  from (\ref{csmaineq}) we deduce that 
\beq
b(x)=-\frac{\Omega^2}{2}\Big((x^1)^2-(x^2)^2\Big)\,\epsilon\,
-b_ix^i+h
\aand
F^i(x)=\omega x^i+f^i\,,\; f^i=\const.
\label{BrdbF}
\eeq
In conclusion, the most general chrono-projective vectorfield for the Brdi\v{c}ka metric has 7 parameters~: 6 isometries and one conformal generator (namely the homothety). In BJR coordinates it is,
\beqa
Y_{B}&&  = \epsilon \Big(\p_u + 
\Omega\big(x^1\tan(\Omega u)\partial_1 - x^2\tanh(\Omega u)\partial_2\big) -\frac{\Omega^2}{2}\big((x^1)^2 - (x^2)^2\big)\p_v\Big) 
\\[6pt]
&&
\,+f^i\p_i+
h\partial_v +\Big(\dfrac{1}{\Omega}\big(b_1\tan(\Omega u)\partial_1
+ b_2 \tanh (\Omega u) \partial_2\big)-b_ix^i\partial_v\Big)
 + \omega\, \big(x^i\p_i+2v\partial_v\big).\quad
\nn
\label{BrdgenCP}
\eeqa
The conformal resp. chrono-projective  factors are
$ 
\omega = - \psi/2 = \const 
$ ;  
$f^i$ and $h$ generate space and vertical translations, respectively, and the $b_i$ generate boosts. 
%%%%%%%%%%%%%%%%%%%%%%%%%%%%%%%%%%%%%%%
%\kikezd{screw for Brdi\v{c}ka}	
%%%%%%%%%%%%%%%%%%%%%%%%%%%%%%%%%%%%%%%

The parameter $\epsilon\in\IR$ generates the  additional  isometry  induced by  $u$-translations.
Expressed in Brinkmann coordinates  the ``screw-charge'' that this isometry generates, 
\beq
\cQ_{\epsilon}^{B}=
p_u +\Omega\Big(\tan(\Omega u)x^1p_1 - \tanh(\Omega u)x^2p_2 \Big)\,
-\frac{\Omega^2}{2}\Big((x^1)^2 - (x^2)^2\Big)\,
\label{BrdCC}
\eeq
turns out to be $P_U$, (minus) the ``Brinkmann'' energy, as expected.

%%%%%%%%%%%%%%%%%%%%%%%%%%%%%%%%%%%%%%%%%%%%%%%%%
\subsection{``Screw'' for circularly polarized periodic (CPP) waves}\label{CPPSec}
%%%%%%%%%%%%%%%%%%%%%%%%%%%%%%%%%%%%%%%%%%%%%%%%%
 
\emph{Circularly polarized periodic waves} (\ref{Bprofile}) with profile \footnote{See also class 14 with $l=0$ in Table 4 of \cite{KeaTu04}.} 
\beq
{\cA} (U) = A_0\cos(\omega U),
\qquad
\mathcal{A}_{\times}(U) = A_0\sin(\omega U),
\qquad
A_0=\const
\label{PerProf}
\eeq
have, beyond the homothety and the usual 5 isometries also a 6th, ``screw'' isometry, obtained by combining broken rotations with broken $U$-translations \cite{exactsol,Carroll4GW,POLPER,Ilderton},
\beq
Y_{CPP}^{scr}=\partial_U  + \frac{\omega}{2} (X^1 \partial_2 - X^2\partial_1)\,.
\label{CPPscrew}
\eeq
We find instructive to outline how this result is recovered using our framework. (We choose $\omega=2$ for simplicity). 
After bringing the system to a $U$-independent form by a suitable rotation, (eqn. \#(5.7) of \cite{POLPER}) we solve the Sturm-Liouville equation (\ref{SLP}) for the profile (\ref{PerProf}) as
\beq
P(u) = 
\begin{pmatrix}
{\cos\Omega_-u}\qquad& \frac{\sin\Omega_+u}{\Omega_+} 
\\[4pt]
-\frac{\sin\Omega_-u}{\Omega_-}\qquad & {\cos\Omega_+u}
\end{pmatrix}\where 
\Omega_\pm^2 = 1 \pm \frac{A_0}{2}\,.
\label{CPPSLP}
\eeq
Then (\ref{BtoBJR}) yields  (\ref{BJRmetric})  with
transverse   metric
\beq
(a_{ij})= 
\begin{pmatrix}
\cos^2\Omega_- u + \frac{\sin^2\Omega_- u}{\Omega_-^2}&\frac{\cos\Omega_- u \sin\Omega_+ u}{\Omega_+} -\frac{\cos\Omega_+ u \sin\Omega_- u}{\Omega_-}  
\\[4pt]
\frac{\cos\Omega_- u \sin\Omega_+ u}{\Omega_+} -\frac{\cos\Omega_+ u \sin\Omega_- u}{\Omega_-} & \cos^2\Omega_+ u + \frac{\sin^2\Omega_+ u}{\Omega_+^2}
\end{pmatrix}\,.
\label{CPPa}
\eeq
The Souriau matrix, calculated using (\ref{CPPa}), is
\beqa
&&(H^{ij}) = \displaystyle\frac{2}{A_0
(\det P)}\times
\\[6pt]
&&\begin{pmatrix}
\frac{\cos\Omega_- u \sin\Omega_+ u}{\Omega_+} - \Omega_- \sin\Omega_- u \cos\Omega_+ u & -1 
\\[4pt]
- 1 & \Omega_+ \sin\Omega_+u \cos\Omega_- u - \frac{\sin\Omega_- u\cos\Omega_+ u}{\Omega_-} 
\end{pmatrix}\,\qquad\quad \nn
\eeqa
 where $\det P$ is the determinant of the Sturm-Liouville matrix (\ref{CPPSLP}), 
 \beq
 \det P = \cos\Omega_- u\cos\Omega_+u + 2\frac{\sin\Omega_+ u \sin\Omega_- u}{\sqrt{4-A_0^2}}.
 \eeq 
A tedious calculation for (\ref{csmaineq}) then yields the sixth isometry in BJR coordinates as 
\beq
Y_{CPP}^{scr}= \partial_u 
-\frac{A_0}{2} \Big((H^{11} x^1- H^{12}x^2 )\partial_1 + (H^{21} x^1 - H^{22}x^2 )\partial_2 \Big) + \frac{A_0}{4} 
\Big((x^1)^2 - (x^2)^2\Big)\partial_v\,,
\label{screwBJR}
\eeq
Circularly polarized periodic waves have thus 6 isometries and one conformal transformation, namely the homothety.

\begin{figure}[h]
\hskip-7mm
\includegraphics[scale=.145]{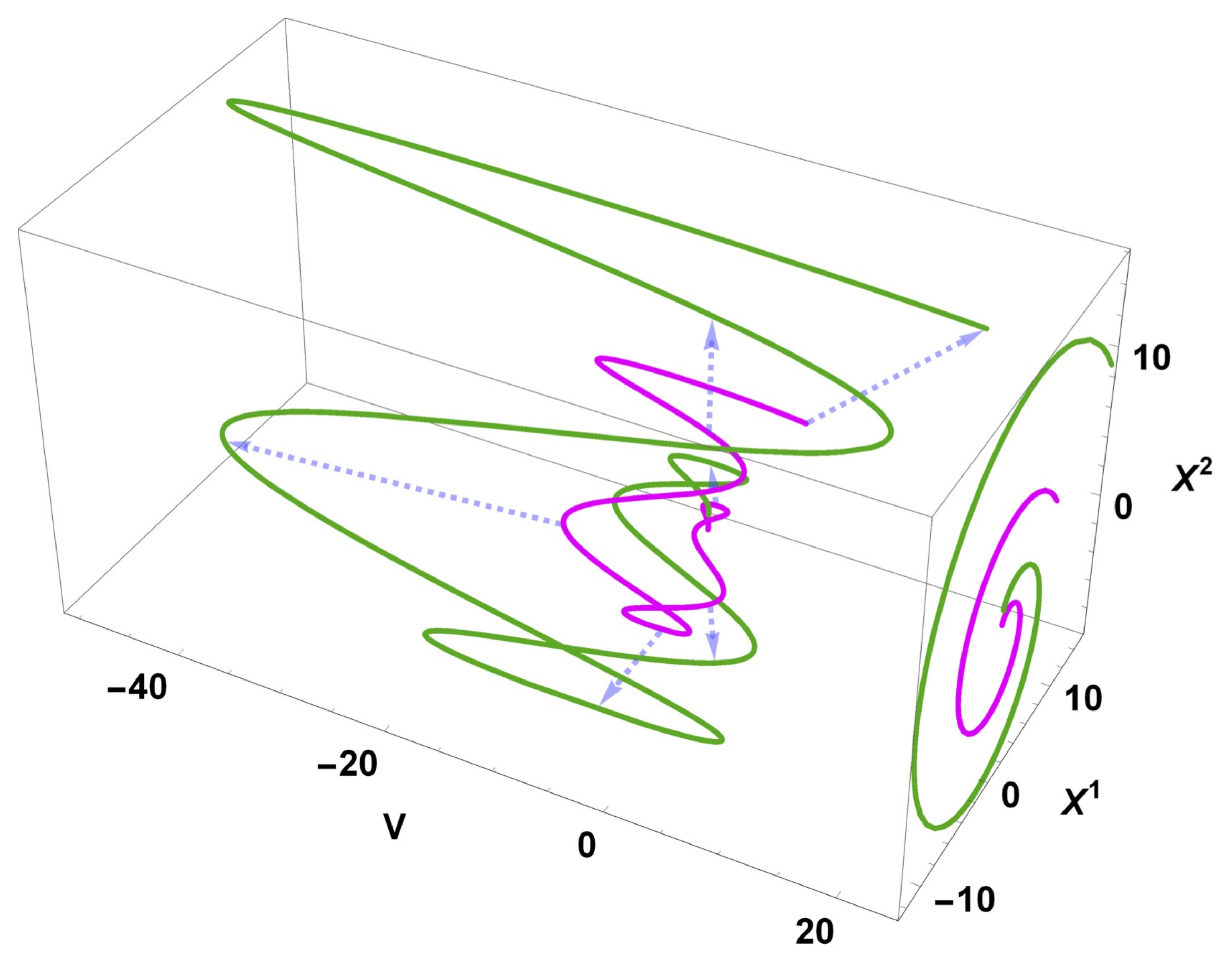}\,
\null\vskip-4mm
\caption{\textit{\small   For the circularly polarized periodic profile (\ref{Bprofile}) with ${\cA}(U) = \cos(U),\,
{\cal A}_{\times}(U) = \sin(U)$  the \blue{\bf homothety} (\ref{homothety0}) takes the trajectory with initial condition $(U_0,\bX_0,V_0)$ [in \magenta{\bf magenta}] into that  with initial condition $(U_0,\chi\,\bX_0,\chi^2\,V_0)$ [in \dgreen{\bf green}].   
\label{blowup}
}}
\end{figure}

\begin{figure}[h]
\hskip-7mm
\includegraphics[scale=.145]{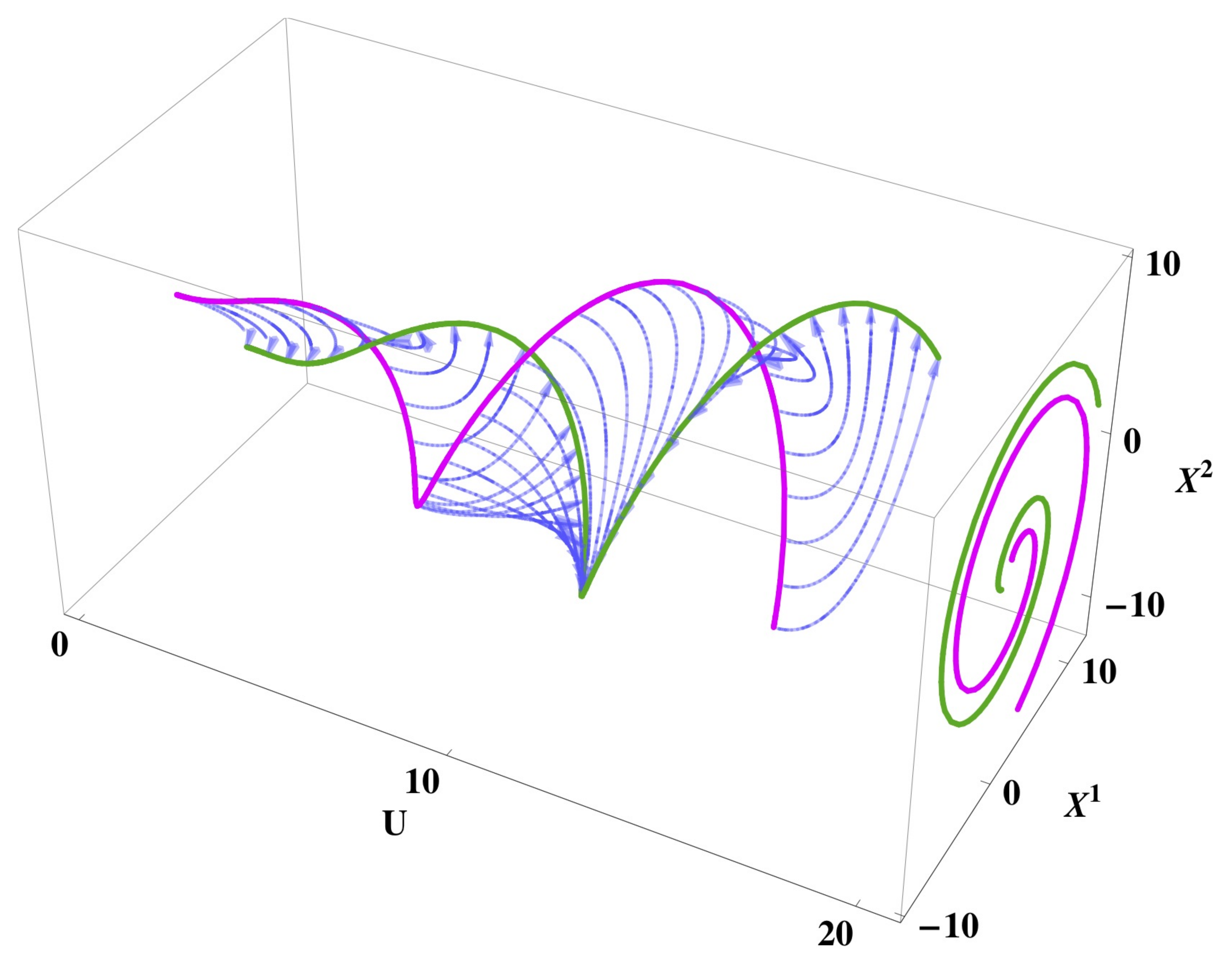} 
\null\vskip-4mm
\caption{\textit{\small Dropping the $V$-coordinate and unfolding the transverse  CPP trajectory by adding  $U$ yields spirals.  
The screw-transformation (\ref{CPPscrew}) (in \blue{\bf blue}) carries the trajectory in \magenta{\bf magenta}  into another trajectory (in \dgreen{\bf green}).
\label{CPPscrewfig}
}}
\end{figure}

Figs.\ref{blowup} and \ref{CPPscrewfig} show how 
 trajectories are taken into trajectories by the homothety and by the screw transformation, respectively.

%%%%%%%%%%%%%%%%%
%%%%%%%%%%%%%%%%%

%%%%%%%%%%%%%%%%%%%%%%%%%%%%%%%%%%%%%%%%%%%%%%%%
\subsection{``Screw'' with expansion}\label{APSec}
%%%%%%%%%%%%%%%%%%%%%%%%%%%%%%%%%%%%%%%%%%%%%%%%

In \cite{AndrPrenc,AndrPrenc2} Andrzejewski and Prencel investigate the memory effect for the linearly polarized gravitational wave with regular $U$-dependent profile \cite{Sippel, Eardley, MaMa, Keane} \footnote{See also class 11ii in Table 4 of \cite{KeaTu04}.} 
\beq
K_{ij}(U) = \frac{\epsilon^2}{(U^2+\epsilon^2)^2}\,\diag(1,-1)\,,
\label{APprof}
\eeq
whose explicit $U$-dependence breaks the  $U$-translation symmetry. 
On the other hand shows that
rescalings are broken with the exception of the homothety. However combining the broken $U$-translation with a broken Schr\"odinger expansion (\ref{expansionM}),
\beq
{Y}^{scr}=Y_{K} +\epsilon^2 \p_U,
\label{screxpvf}
\eeq
we to call here a ``screwed expansion'',  generates a conformal transformation
\beq
L_{Y^{scr}} g_{\mu\nu} = 2U g_{\mu\nu}\,, 
\qquad
L_{Y^{scr}}\xi= 0\,. 
\eeq
%%%%%
The conserved quantity associated with \eqref{screxpvf} 
\beq
\cQ^{scr} = (U^2 + \epsilon^2)P_U + U X^i P_i - \frac{\bm{X}^2}{2}P_V
\eeq
satisfies
\beq
\{\cQ^{scr},\cH\} = 2 U \cH \,
\eeq
and is therefore conserved for null geodesics. 

In conclusion, the Brinkmann metric with profile (\ref{APprof}) provides us with an
 example with 5 isometries and \emph{two} conformal generators, namely the homothety and the ``screw" (\ref{screxpvf}).
  
 \goodbreak

%%%%%%%%%%%%%%%%%%%%%%%%%%%%%%%%
\subsection{``Screw'' with U-V boost}\label{IlderSec}
%%%%%%%%%%%%%%%%%%%%%%%%%%%%%%%%

Ilderton  \cite{Ilderton,IldertonPC}  
 mentions that for the [singular] profile
\beq
K_{ij}(U)=
\frac{K^0_{ij}}{(1+U)^2},
\qquad K^0_{ij}=\const
\label{Ilderprof}
\eeq
the manifest breaking of $U$-translation invariance can be cured by ``screw-combining'' it with a (broken) boost (\ref{UVboostM}), i.e., $
Y_{UV}^{scr}=Y_U+Y_{UV}.
$
His statement is confirmed by calculating the Poisson bracket of the associated charge with the Hamiltonian,
\beq
\cQ_{UV}^{scr}=P_U + \cQ_{UV}\,,
\qquad
\big\{\cQ_{UV}^{scr},\cH\big\}=0.
\eeq

The conformal (resp. chrono-projective) factors are
$\omega= 0$ and $\psi= 1\,.$
Adding the homothety, we  end up with a chrono-isometry plus a chrono-conformal transformation in addition to the standard 5 isometries.
Another way of understanding  this is to observe that  $U\to U-1$ carries the profile (\ref{Ilderprof}) to the form (\ref{inverseU2}), whose U-V boost symmetry was established in sec. \ref{homoSec}.

%%%%%%%%%%%%%%%%%%%%%%%%%%%%%%%%%%%%
\section{Conclusion}\label{Concl}
%%%%%%%%%%%%%%%%%%%%%%%%%%%%%%%%%%%%

Plane gravitational waves have long been known to admit, generically, a 5-parameter isometry group \cite{BoPiRo,EhlersKundt,Sou73,Torre,exactsol}.
The homothety (\ref{homothety0}) is a universal conformal generator.
For a non-conformally-flat spacetime, the maximum number of conformal Killing vectors is $7$ \cite{Sippel, Eardley, MaMa, Keane,exactsol,HallSteele,KeaTu04}. 
The associated conserved quantities  determine the transverse-space trajectory $\bX(\sigma)$ in (\ref{ABXeq}) \cite{Sou73,Carroll4GW} and for null geodesics $\cQ_{hom}$ in (\ref{Qhomot}) allows us to find also the vertical motion according to  (\ref{VUGW}).

The homothety is a chrono-projective transformation  introduced by Duval et al. \cite{DThese,5Chrono}. The fundamental importance of the latter becomes clear from two, related contexts.

Firstly, the chrono-projective property  (\ref{Chronocond}) is precisely what we need to derive conserved quantities for null geodesics in the gravitational wave spacetime and -- unexpectedly -- also in the underlying non-relativistic dynamics \cite{KHarmonies}. The  conserved quantities (\ref{Qdown}) they generate involve a novel term, namely the  action integral of the underlying non-relativistic dynamics.

Secondly, for an exact plane gravitational wave, \emph{all} conformal vectors are 
chrono-projective \cite{MaMa}. Using the chrono-projective condition makes it simpler to determine the conformal transformations for a given profile, as illustrated in sec. \ref{Examples}. Using BJR coordinates is particularly convenient.

Since we have not made use of Ricci flatness in our calculations in sect.\ref{BJRSec}, our solutions  (\ref{csKvs}) should apply for the special conformal Killing vectors of any pp-wave.

Actually,  for a type-N (non-flat) null fluid spacetime \cite{KeaTu04} arbitrary conformal vector fields satisfy the chrono-projective conditions (\ref{confodef0}) and (\ref{Chronocond}). Although different coordinates were used, one can notice the similarity between their  eqns. $\#$ (19) and our (\ref{csKvs}). However, unlike in their case, the form of the unknown functions $F(x)$ and $b(x)$ is readily determined, justifying our preference for BJR coordinates.

As it is illustrated in sect. \ref{Examples}, the isometry group of gravitational  waves can, in special instances, be enlarged to 6 parameters \cite{exactsol, Sippel, Eardley, MaMa, Keane}. 
 This is plainly the case when the profile is $U$-independent so that $U$-translations are isometries; an example is given by the Brdi\v{c}ka metric (\ref{Brdmetric}). 
When the profile does depend on $U$, $U$-translations are manifestly broken, however they can, under special circumstances, be combined with another broken symmetry generator yielding an additional ``screw-type'' conformal symmetry \cite{exactsol,POLPER,Ilderton}. Examples are presented in (\ref{inverseU2}) and in sec.\ref{Examples}.  

Chrono-projective transformations may extend to Newton-Cartan framework \cite{DuvalNC, NewtonCartan} as conformal extensions of Carroll manifolds. 

At last, we mention that our investigations have some overlap with those in \cite{Morand:2018tke,Igata}, as we discovered during the final phases of this research.

\begin{acknowledgments} 
We are grateful to Christian Duval (1947-2018) for his advices during 40 years of friendship and collaboration, and his contribution at the early stages of this project. This paper is dedicated to his memory. We would  like to acknowledge Gary Gibbons for  discussions during the long preparation of this paper. We thank also X. Bekaert, P. Pio Kosi\'nski, Krzysztof Andrzejewski \cite{PKKA} Anton Ilderton \cite{IldertonPC} and Nikolaos Dimakis  \cite{DimakisPC} and Thomas Helpin. 
 ME thanks the \emph{Institute of Modern Physics} of the Chinese Academy of Sciences in Lanzhou and  the \emph{Denis Poisson Institute of Orl\'eans-Tours University}.
PH  thanks the \emph{Institute of Modern Physics} of the Chinese Academy of Sciences in Lanzhou for hospitality. This work was partially supported by the Chinese Academy of Sciences President's International Fellowship Initiative (No. 2017PM0045), and by the National Natural Science Foundation of China (Grant No. 11975320).
\end{acknowledgments}
\goodbreak

%%%%%%%%%%%%%%%%%%%%%%%%%%%%%%%

\end{document}